\documentclass[a4paper,11pt]{article}
\usepackage{jheppub} 

\usepackage[T1]{fontenc} 
\usepackage{xcolor}
\usepackage{caption}
\usepackage{subcaption}
\usepackage[normalem]{ulem}

\title{\boldmath Krylov Complexity for Jacobi Coherent States}

\author[a,b]{S. Shajidul Haque,}
\author[a,b]{Jeff Murugan,}
\author[a]{Mpho Tladi,}
\author[a,b]{and Hendrik J.R. Van Zyl\,}

\affiliation[a]{The Laboratory for Quantum Gravity \& Strings, Department of Mathematics \& Applied Mathematics, University of Cape Town, Cape Town, South Africa}
\affiliation[b]{The National Institute for Theoretical and Computational Sciences, Private Bag X1, Matieland, South Africa}

\abstract{We develop computational tools necessary to extend the application of Krylov complexity beyond the simple Hamiltonian systems considered thus far in the literature. As a first step toward this broader goal, we show how the Lanczos algorithm that iteratively generates the Krylov basis can be augmented to treat coherent states associated with the Jacobi group, the semi-direct product of the 3-dimensional real Heisenberg-Weyl group $H_{1}$, and the symplectic group, $Sp(2,\mathbb{R})\simeq SU(1,1)$. Such coherent states are physically realized as squeezed states in, for example, quantum optics. With the Krylov basis for both the $SU(1,1)$ and Heisenberg-Weyl groups being well understood, their semi-direct product is also partially analytically tractable. We exploit this to benchmark a scheme to numerically compute the Lanczos coefficients which, in principle, generalizes to the more general Jacobi group $H_{n}\rtimes Sp(2n,\mathbb{R})$.}

\begin{document} 
\maketitle
\flushbottom

\bibliographystyle{JHEP}
\section{Introduction}

Quantum complexity \cite{nielsen00, Watrous_2008} aims to quantify how difficult it is to prepare a desired final state {\color{black}{of a quantum system,} given some initial configuration. This requires information about the initial state, the set of unitary operations at one's disposal} and a measure of the difficulty in performing each of the unitary operations.  As such, the definition of quantum complexity is not unique and there are, by now, several candidates that are the subject of much active research (see e.g. reviews \cite{Chapman:2021jbh,Chen:2021lnq} and references therein).  \\ 

\noindent 
One such measure, the so-called {\it Krylov-} or {\it spread-complexity} \cite{Parker:2018yvk}, takes as a starting point the time-evolution of a ``simple'' reference state in the Hilbert space. This is intuitively appealing since the time-evolution will naturally involve the spectrum of the system under consideration which encodes many of the properties of the system in an invariant way. For example, in chaotic systems, the spectral statistics of the Hamiltonian are conjectured to asymptote to those of a random matrix ensemble at late times \cite{Bohigas:1983er} .  As such, it is reasonable to expect that Krylov complexity may be used as a diagnostic of quantum chaotic systems, to supplement other diagnostics \cite{Ali:2019zcj,Bhattacharyya:2019txx,Bhattacharyya:2020art,Bhattacharyya:2020iic}.  Furthermore, the space of states explored by integrable and chaotic systems differ dramatically \cite{Balasubramanian:2021mxo}, with the latter expected to explore a Hilbert space subspace in an ergodic way.  To quantify the complexity of the time-evolved state one may imagine decomposing it into a linear combination of states from some basis, where the basis vectors are ordered in increasing complexity.  The Krylov basis minimises the complexity of the time-evolved state over these choices of basis \cite{Caputa:2021sib, Balasubramanian:2022tpr}.  \\ \\
Spread complexity has already been shown to be sensitive to topological phase transitions \cite{Caputa:2022eye, Caputa:2022yju} as exemplified in the Su-Shriefer-Heeger (SSH) model {\footnote{Previously, the circuit complexity for this model was studied in \cite {Ali:2018aon}.} of polyacetylene \cite{Su:1979ua} polymer chains as well as the 1-dimensional Kitaev chain \cite{Kitaev_2001} p-wave superconductor,  a feature that is robust against a particular choice of reference state \cite{Caputa:2022yju}. Key to the computation of the spread complexity in these quantum systems was the fact that in each case, the Hamiltonian can be expressed in terms of $SU(2)$ coherent states, for which exact analytic results relating to spread complexity are known\footnote{ See also \cite{Muck:2022xfc} where analytic results are obtained from an input of simple Lanczos coefficients.} \cite{Caputa:2021sib}. Still, as remarkable as both these models are, they are a set of measure zero in the space of novel quantum systems. Ideally, we would like to understand the nature and behaviour of Krylov complexity in more general systems, with the 
Hamiltonian selected from a larger symmetry group, or even where perturbative treatments may be unavoidable. This article aims to initiate development in just this \footnote{ Note that modifications of the Lanczos algorithm has been developed in the context of diagonalizing sparse matrices (see e.g. \cite{Parlett1979TheLA, Simon1984TheLA, Gambolati}).  Here we are interested in a modification aimed at the computations of spread complexity of quantum states. }. Specifically, we will focus on the {\it Jacobi algebra}, a semi-direct sum of $su(1,1)$ and the Heisenberg algebra \cite{Berceanu:2007zz}.\\

\noindent
 Krylov complexity for the Jacobi group is a conservative, yet interesting extension of the $SU(2)$ case. This is in equal measure because the coherent states attached to the Jacobi group have an actual physical realization as squeezed states in quantum optics \cite{Berceanu2009TheJG} and, more importantly, since exact analytic expressions may be obtained for both $SU(1,1)$ and Heisenberg Krylov complexity \cite{Caputa:2021sib}, there is hope that similar expressions for their semi-direct product exist also. Indeed tensor products of exactly solvable algebras have also been studied \cite{Kressner} and, in fact, do yield analytic expressions (see e.g. \cite{Adhikari2022CosmologicalKC, Caputa:2022eye, Caputa:2022yju}). Crucially however, in the case of the Jacobi algebra, the non-vanishing commutation relations between the elements of the $su(1,1)$ and Heisenberg algebras give rise to involved combinatorics in determining the Krylov basis. In this sense, the Jacobi algebra is a simple prototype for more general Lie algebras.     \\ \\
Conventionally, the Krylov basis is generated by the Lanczos algortihm \cite{Lanczos1950AnIM} (see also the review \cite{viswanath2008recursion}), an iterative algorithm that generalizes the Gram-Schmidt procedure for constructing an orthonormal basis. However, a naive application of the Lanczos algorithm to the Jacobi group fails to efficiently compute the Lanczos coefficients and must be modified. The purpose of this article is to propose such a modification. Our proposed algorithm generates a family of vectors each step in the iterative scheme, converging on the Krylov basis and in which at the $k{th}$ step contains the first $k$ Krylov vectors.  Explicit computations in this modified recursion method may be performed using differential operators acting on Jacobi coherent states \cite{Perelomov}.  Moreover, by terminating the procedure at any finite order one obtains an approximate Krylov basis.  We demonstrate that this approximation scheme is accurate for a range of parameter choices.  \\ \\
This article is structured as follows: In section 2 we review the idea of Krylov complexity in the context of symmetry groups, paying particular attention to the cases of $SU(1,1)$ and the Heisenberg group $H_{1}$. This is followed by a detailed discussion of our proposed modification of the Lanczos algorithm in section 3 and an explicit computation of the Lanczos coefficients for the Jacobi group as well as the associate K-complexity in section 4. We conclude with some discussion and speculation of future directions in section 5.

\section{Krylov Complexity}
\label{KrylovSection}

 We begin with a brief overview of idea of K-complexity for quantum states. For the computation of state $K$-complexity we require a reference state (chosen from a Hilbert space with an appropriate inner product) as well as a Hamiltonian acting on this reference state.  The starting point for the analysis is the time-evolved reference state
\begin{equation}
    |t\rangle \equiv e^{i t H} |\phi_0\rangle\,,
 \end{equation}
By expanding this in $t$, one obtains a basis for the space of accessible target states
\begin{equation}
    |O_n) \equiv H^n |\phi_0\rangle \,,
\end{equation}
However, this basis is not, in general, orthogonal with respect to to the Hilbert space inner product.  The Krylov basis is obtained by applying a Gram-Schmidt orthogonalisation procedure on the basis $\{|O_n)\}$.  One popular method by which to achieve this is the Lanczos algorithm, or recursion method by which we define,
\begin{eqnarray}
    |K_0\rangle & \equiv & |\phi_0\rangle\,, \nonumber \\
    |A_{n+1}) & = & (H - a_n) |K_n\rangle - b_n |K_{n-1}\rangle\,, \nonumber \\
    |K_n\rangle & \equiv &  (A_n| A_n)^{-\frac{1}{2}}  |A_n)   \,, \nonumber
\end{eqnarray}
and where 
\begin{eqnarray}
a_n & = & \langle K_n | H |K_n \rangle\,, \nonumber \\
b_n & = & (A_n| A_n)^{ \frac{1}{2} } \,.
\end{eqnarray}
A notable feature of this construction is that it tri-diagonalizes the Hamiltonian, in the sense that in the Krylov basis,
\begin{equation}
    H |K_n\rangle = a_n |K_n\rangle + b_n |K_{n-1}\rangle + b_{n+1} |K_{n+1}\rangle\,.
\end{equation}
There are two ways of interpreting this: i) knowledge of the Krylov basis determines the Lanczos coefficients as Hamiltonian matrix elements and ii) the Lanczos coefficients give the Krylov basis by means of the recursive Lanczos algorithm.  \\ \\
Finally, given the Krylov basis, the spread complexity (or state K-complexity) of a target state $|\psi\rangle$ is given by
\begin{equation}
    K = \sum_{n=0}^\infty n  \langle \psi | K_n \rangle  \langle K_n| \psi\rangle\,.
\end{equation}
This algorithm may be applied, in principle, to any Hamiltonian, however when the Hamiltonian itself is an element of some symmetry algebra, the computation simplifies considerably, in large part due to the utility of coherent states attached to the symmetry \cite{Perelomov}.

\subsection{$SU(1,1)$ and Heisenberg K-complexity}
\label{KExampleSection}

Our primary interest in this article is in the study of Krylov complexity when the Hamiltonian is expressed in terms of elements of the Jacobi algebra.  Physically, this is the most general choice of Hamiltonian up to quadratic order in the position and momentum operators. As the semi-direct sum of $su(1,1)$ and the Heisenberg algebra, the Jacobi algebra is generated by the operators $L_{\pm}, L_{0}, l_{\pm}$ and $l_{0}$ satisfying the commutation relations
\begin{eqnarray}
\left[ L_0, L_{\pm} \right] = \pm L_{\pm} \ \ \  & & \ \ \  \left[ L_{-}, L_{+}\right] = 2 L_0 \nonumber \\
\left[ l_{-}, l_{+}\right] = l_0 \ \ \  & & \ \ \ \left[ l_0, l_{\pm}\right] = 0
\end{eqnarray}
respectively, as well as the 
``cross-commutators''
\begin{eqnarray}
\label{Jacobi-cross-comm}
\left[ L_0, l_{\pm} \right] & = & \pm \frac{1}{2} l_{\pm}   \nonumber \\
\left[ L_{+}, l_{-} \right] & = & -l_{+} \\
\left[ L_{-}, l_{+}\right] & = & l_{-}  \nonumber
\end{eqnarray}
with all other commutators vanishing. It is this final set of commutators, of course, that enlarges the Jacobi algebra from the direct sum of $su(1,1)$ and Heisenberg algebra.  \\ \\
Since exact analytical results for the Lanczos coefficients and Krylov complexity for Hamiltonians selected from $su(1,1)$ or the Heisenberg algebra may be  readily obtained (see for example \cite{Caputa:2021sib} and references therein), one might be forgiven for expecting that the Krylov basis (with say, the lowest weight reference state) for a Jacobi algebra Hamiltonian may also be computed exactly, and in closed form. However, as we will demonstrate, the additional commutators in \eqref{Jacobi-cross-comm} 
result in rather involved combinatorics associated with the Gram-Schmidt procedure generating the Krylov basis.\\

\noindent
Before proceeding to the specifics of the Jacobi K-complexity, let's first summarise the simpler cases of $SU(1,1)$ and Heisenberg K-complexity in a form that will allow for necessary modifications. To this end, it will be particularly useful to phrase the computation in the language of coherent states and differential operators, which will generalise nicely when we consider the Jacobi group K-complexity.  

\subsubsection{$SU(1,1)$}
Let's begin with the more conceptually familiar $su(1,1)$ case. The most general $su(1,1)$ Hamiltonian can be written as
\begin{equation}
    H = \alpha L_{+} + \alpha^* L_{-} + \gamma L_0\,,   
\end{equation}
with (complex) coefficients $\alpha$ and $\gamma$. As reference state we choose the lowest weight, $|h\rangle$, defined by
\begin{eqnarray}
L_0 |h\rangle  =  h |h\rangle\,, \quad
L_{-} |h\rangle = 0\,.    
\end{eqnarray}
The space of states spanned by $\left\{ H^m |h\rangle \ , \ m=0,1,\cdots n \right\} $ is then equivalent the space spanned by $\left\{ L_{+}^m|h\rangle \ , \ m=0,1,\cdots n   \right\}$ for each value of $n$.  Indeed, provided $\alpha \neq 0$, the Krylov basis is given by 
\begin{equation}
    |K_n\rangle = \left(\langle h | L_{-}^n L_{+}^n |h 
    \rangle\right)^{-\frac{1}{2}} L_{+}^n |h\rangle\,.
\end{equation}
Note that this is independent of the coefficients parametrising the Hamiltonian.  
Consequently, the (unnormalised) $SU(1,1)$ Perelemov coherent states
\begin{equation}
    |z) = e^{z L_{+}} |h\rangle = \sum_{n=0}^{\infty} \frac{z^n}{n!} L_{+}^n |h\rangle\,,
\end{equation}
provide a natural basis in which quantities associated with Krylov complexity may be computed. For what follows, a particularly convenient representation for the $su(1,1)$ generators is as differential operators, which act on the associated coherent states as 
\begin{eqnarray}
L_{+} |z) & = & \partial_z |z)\,,   \nonumber \\
L_0 |z) & = & (h + z \partial_z) |z)\,,  \\
L_{-} |z) & = & (2 z h + z^2 \partial_z) |z)\,.   \nonumber 
\end{eqnarray}
Two further ingredients for the Lanczos algorithm are required; the overlap between coherent states,
\begin{equation}
    (\bar{z}|z) = \sum_{n} \frac{z^n \bar{z}^n}{(n!)^2} \langle h | L_{-}^n L_{+}^n |h\rangle  =  (1 - z \bar{z})^{-2h} 
\end{equation}
and the coherent-state expectation value of the Hamiltonian, 
\begin{eqnarray}
     (\bar{z}|H |z) &=&  h \left( 2 z \alpha^* + 2 \bar{z} \alpha + (1 + z \bar{z}) \gamma    \right)(1 - z \bar{z})^{-(2h+1)}   \nonumber \\
     & = & \sum_{n=1}^\infty \frac{\bar{z}^{n} z^{n-1}}{(n-1)! n!} \sqrt{  \langle h| L_{-}^{n-1} L_{+}^{n-1} |h\rangle \langle h| L_{-}^{n} L_{+}^{n} |h\rangle } \ b_n  \nonumber \\
     &+& \sum_{n=0}^\infty \frac{\bar{z}^{n} z^{n+1}}{(n+1)! n!} \sqrt{  \langle h| L_{-}^{n+1} L_{+}^{n+1} |h\rangle \langle h| L_{-}^{n} L_{+}^{n} |h\rangle } \ b_{n+1} \nonumber \\
     &+&  \sum_{n=0}^\infty \frac{\bar{z}^{n} z^{n}}{(n!)^2 }   \langle h| L_{-}^{n} L_{+}^{n} |h\rangle a_n\,.
\end{eqnarray}
Putting this together allows us to  extract the Lanczos $a_{n}$ and $b_{n}$ coefficients as
\begin{eqnarray}
b_n & = & \left. \hat{O}_n^\dag \hat{O}_{n-1}   \left(\bar{z}|H |z\right) \right|_{z, \bar{z} \rightarrow 0}\,,    \nonumber \\
a_n & = & \left. \hat{O}_n^\dag \hat{O}_{n}
(\bar{z}|H |z) \right|_{z, \bar{z} \rightarrow 0}\,, 
\end{eqnarray}
where the family of differential operators,
\begin{equation}
\hat{O}_{n} = \frac{1}{\sqrt{\langle h| L_{-}^n L_{+}^n |h\rangle } } \partial_z^n\,.
\end{equation}

\subsubsection{Heisenberg}

Moving on, similar results may be obtained for the Heisenberg algebra. Here the most general hermitian Hamiltonian operator we may write down (dropping the contribution proportional to the identity) is 
\begin{equation}
    H = \beta l_{+} + \beta^* l_{-}\,.
\end{equation}
Again, a natural choice of reference state is the lowest weight state, $|0\rangle$ satisfying 
\begin{equation}
l_{-}|0\rangle = 0\,.
\end{equation}
 The space of states spanned by $\left\{ 
 H^m|0\rangle \ , \ m=0, 1, \cdots n\right\}$ is equivalent to the space spanned by $\left\{ 
 l_{+}^m |0\rangle \ , \ m=0, 1, \cdots n\right\}$.  For any non-zero value of $\beta$, the recursion method described above yeilds the  Krylov basis
\begin{equation}
    |K_n\rangle = \left( \langle 0| l_{-}^n l_{+}^n |0 \rangle \right)^{-\frac{1}{2}} l_{+}^n |0\rangle\,. 
\end{equation}
Coherent states in this case are the so-called Glauber coherent states \cite{Glauber1963CoherentAI},
\begin{equation}
    |z) = e^{z l_{+}} |0\rangle\,. 
\end{equation}
As before, in order to extract the Lanczos coefficients, we will require the expectation value of the Hamiltonian in the coherent state basis. This is simply computed as,
\begin{equation}
    ( \bar{z} | H |z ) = (\alpha \bar{z} + z \alpha^*) e^{z \bar{z}}\,,
\end{equation}
and along with the family of differential operators
\begin{equation}
    \hat{O}_n = \frac{1}{\sqrt{\langle 0 | l_{-}^n l_{+}^n |0\rangle } } \partial_{z}^n\,,
\end{equation}
provides a simple and exact expression for the Lanczos coefficients 
\begin{equation}
    b_n = \hat{O}_n^\dag \hat{O}_{n-1}  ( \bar{z} | H |z )\,.
\end{equation}
Given the simplicity of these expressions, as well as the (relative) ease with which they can be derived, one may be justified in hoping that similar expressions can be derived for Hamiltonian systems that combine these two algebras in some way.

\subsection{Jacobi group}

We now turn our attention to the Jacobi group and describe the details of our preferredrepresentation. As a first observation, note that the operators $L_0$, $l_0$ and the Casimir $C= (L_0)^2 - \frac{1}{2}(L_{+} L_{-} + L_{-} L_{+})$ may be simultaneously diagonalized.   The simultaneous eigenstate has eigenvalues
\begin{eqnarray}
L_0 |h, n; M\rangle & = & (h + n)|h, n; M\rangle\,,\\
l_0 |h, n; M\rangle & = & M|h, n; M\rangle\,.\nonumber   \label{diagElements}
\end{eqnarray}
The ladder operators act on this state as
\begin{eqnarray}
L_{+} |h, n; M\rangle & = & \sqrt{(n+1)(2 h + n)} |h, n+1; M\rangle\,, \nonumber \\
l_{+} |h, n; M\rangle & = & \sqrt{M}\sqrt{2n+1} |h, n+\frac{1}{2}; M\rangle\,,   \nonumber \\
L_{-} |h, n; M\rangle & = &  \sqrt{n(2 h + n -1)} |h, n-1; M\rangle\,, \nonumber \\
l_{-} |h, n; M\rangle & = & \sqrt{M} \sqrt{2n} |h, n-\frac{1}{2}; M\rangle\,.   \label{ladderElements}
\end{eqnarray}
In addition, there is one more consistency condition that needs to be imposed. It isbest seen by studying the overlap
\begin{equation}
    \langle h, 0; M | L_{-} l_{+} l_{+}| h, 0; M \rangle  =  \sqrt{4h} M = M \ \ \ \Rightarrow \ \ \  h = \frac{1}{4} \ \ \ \textnormal{or} \ \ \ M = 0 \,.   \label{hMCondition} 
\end{equation}
We will be interested in the case $M \neq 0$ and will therefore take $h = \frac{1}{4}$.  \\ \\
The Perelemov coherent states \cite{Perelomov} is given by \begin{equation} 
|z_1, z_2) =  e^{z_1 l_{+}} e^{z_2 L_{+}} |h,0, M\rangle
\end{equation} with the normalised vectors given by $|z_1, z_2\rangle = (\bar{z}_1, \bar{z}_2| z_1, z_2)^{-\frac{1}{2}} |z_1, z_2)$. A useful quantity for our purposes is the coherent state overlap, given by
\begin{eqnarray}
& & \langle h,0; M | e^{\bar{z}_2 L_{-}} e^{\bar{z}_1 l_{-}} e^{z_1 l_{+}} e^{z_2 L_{+}} |h, 0; M\rangle   \nonumber \\
&=& \langle h,0; M | e^{\frac{z_2}{1 - z_2 \bar{z}_2} L_{+}} e^{\frac{z_1 + \bar{z}_1 z_2}{1 - z_2 \bar{z}_2} l_{+}} e^{-2 \log\left( 1 - z_1 \bar{z}_1\right) L_0} e^{\frac{1}{2} \frac{2 z_1 \bar{z}_1 + \bar{z}_1^2 z_2 + (z_1)^2 \bar{z}_2}{1 - z_2 \bar{z}_2} l_0} e^{\frac{\bar{z}_1 + z_1 \bar{z}_2}{1 - z_2 \bar{z}_2} l_{-}} e^{\frac{\bar{z}_2}{1 - z_2 \bar{z}_2} L_{-}} |h, 0; M\rangle   \nonumber \\
& = & (1 - z_2 \bar{z}_2)^{-2 h} e^{\frac{M}{2} \frac{2 z_1 \bar{z}_1 + \bar{z}_1^2 z_2 + (z_1)^2 \bar{z}_2}{1 - z_2 \bar{z}_2}}\,.    \label{JacobiCSOL}
\end{eqnarray}
In arriving at the final expression in \eqref{JacobiCSOL} we have made extensive use of the BCH formula.  The operator actions (\ref{ladderElements}) may be verified using the overlap by taking appropriate derivatives.  Finally, note that the factor of $M$ may be absorbed by rescaling $z_1 \rightarrow \frac{z_1}{\sqrt{M}}$.  As such, and especially when obtaining numeric results, we will set $M=1$.

\section{Jacobi Group Krylov basis}

For both the $SU(1,1)$ and Heisenberg symmetry groups a key feature allowing for exact solutions is that the span of the vectors $\left\{ H^m |\phi_0\rangle , \ m=0,1,\cdots n\right\}$ matches up exactly with the span of $\left\{ L_{+}^m |\phi_0\rangle \ , \ m=0,1,\cdots n  \right\}$ and $\left\{ l_{+}^m |\phi_0\rangle \ , \ m=0,1,\cdots n \right\}$ respectively.  This plays a central role in readily identifying the Krylov basis. The Jacobi group however, possesses two sets of ladder operators, one of which increases the eigenvalue of $L_0$ by $1$ while the other by $\frac{1}{2}$, as in (\ref{ladderElements}). After $n$ applications of a general Jacobi Hamiltonian one explores a space consisting of specific linear combinations of the first $n$ half-integer labelled states and the first $(n+1)$ integer labelled states.  Because of this we cannot extract the Krylov basis as easily as in the $SU(1,1)$ or Heisenberg cases.  \\ \\
As is, by now, well appreciated, 
the Lanczos algorithm provides an algorithmic way to generate the Krylov basis vectors. Less well-known though is the fact that when studying problems involving an infinite-dimensional Krylov subspace (especially at late times) the algorithm is far less efficient since it generates the Krylov basis vectors one at a time and a very large number of basis vectors are required in order to accurately describe late-time behavior.  \\ \\
It is possible to circumvent this problem by modifying the Lanczos algorithm with the aid of some intuition for the Krylov basis vectors. In the next subsections we will illustrate this modified iterative method, but first let's point out  
a few features worth highlighting:  After $m$ steps in our iteration the first $m$ Krylov basis vectors are obtained exactly, precisely as one would in the Lanczos algorithm.  Unlike the Lanczos algorithm however, one obtains not a single vector but a basis of vectors simultaneously.  By terminating the scheme at finite order we obtain an approximate Krylov basis.  The Hamiltonian, when expressed in this approximate basis, is tri-diagonal so that the scheme provides a systematic way to approximate the Lanczos coefficients. Alhough our focus in this paper is a Hamiltonian selected from the Jacobi algebra the scheme is not dependent on this fact and should generalise.

\subsection{An iterative scheme}

We now turn our attention to an explicit Hamiltonian built from generators of the Jacobi algebra,
\begin{equation}
    H = \alpha L_{+} + \alpha^* L_{-} + \beta l_{+} + \beta^* l_{-} + \gamma L_0   \label{JacobiH}
\end{equation}
 We will also occasionally refer to the phases of these complex parameters, using $\alpha = |\alpha|e^{i a}$, $\beta = |\beta|e^{i b}$.  As reference state we choose $|h,0,M\rangle$ which provides a large stability subgroup\footnote{Note, however, that we can easily extend the results we obtain to any choice of reference state $U|h,0,M\rangle$ with $U\in SU(1,1)\rtimes H_{1}$.  Such a choice of reference state is equivalent to studying the $|h,0,M\rangle$ reference state with $U H U^\dag$ as Hamiltonian.  This unitary transformation may be absorbed as a coordinate transformation of the constants, $\alpha, \beta$ and $\gamma$.  In appendix \ref{refStateAppendix} we highlight a special choice of reference state for which the system is exactly solvable.}, since it is a lowest weight state of $L_0$, annihilated by both $L_{-}$ and $l_{-}$. 
Our intuition is guided by the up-ladder part of the Hamiltonian
which, after $n$ applications on the reference state reads
\begin{equation}
    (\alpha L_{+} + \beta l_{+})^n |h, 0, M\rangle
\end{equation}
While these vectors are not orthogonal to the space spanned by $\left\{ H^{m} |h, 0, M\rangle \ , \ m=1,2,\cdots, n-1  \right\}$, it is straightforward to identity a component that is; namely
\begin{equation}
    \left|A_{n}^{ (0)}\right\rangle = \frac{\left( \alpha L_{+}^n + n \beta L_{+}^{n-1} l_{+} \right) |h, 0, M\rangle}{\sqrt{ |\alpha|^2\left\langle L_{-}^n L_{+}^n \right\rangle + n^2 |\beta|^2\left\langle L_{-}^{n-1} l_{-} l_{+} L_{+}^{n-1} \right\rangle   }} \label{An0}
\end{equation}
This observation is the starting point for our approximation scheme and represents the ``$0^{\mathrm th}$'' order approximation for the Krylov basis.  This family of vectors is what one may have guessed as the Krylov basis based on the intuition for the $su(1,1)$ and Heisenberg cases.  However, these vectors represent a particular linear combination of $|h,n,M\rangle$ and $|h,n-\frac{1}{2},M\rangle$.  Subsequent actions of the Hamiltonian will mix with the orthogonal component of this vector and we need to determine the correct linear combination describing the evolution.\\ \\
To proceed from this point we thus have to find a family of vectors orthogonal to $|A_n^{(0)}\rangle$ and then find the appropriate linear combination of these vectors.  The first step is easy to perform in this example as we can determine it uniquely
\begin{equation}
    \left|B_{n}^{(1)}\right\rangle = \frac{(\beta^* L_{+}^n - \alpha^* L_{+}^{n-1}l_{+} )| h, 0, M\rangle}{ \sqrt{ |\beta|^2\langle L_{-}^n L_{+}^n \rangle + |\alpha|^2\langle L_{-}^{n-1} l_{-} l_{+} L_{+}^{n-1} \rangle   } }
\end{equation}
The second requires finding the appropriate linear combination
\begin{equation}
    \left|A_n^{(1)} \right\rangle = \frac{1}{\sqrt{1 + |N_{1,n}|^2 }}\left( |A_{n}^{(0)} \rangle + N_{1,n} |B_{n-1}^{(1)}\rangle \right)   \label{firstOrder}
\end{equation}
The coefficients $N_{1,n}$ can be determined as
\begin{equation}
    N_{1,n} = \frac{ \left\langle B_{n-1}^{(1)}  | H^n| h, 0, M  \right\rangle   } { \left\langle A_n^{(0)} | H^n| h, 0, M \right\rangle  }  \label{firstOrderCoeff}
\end{equation}
The formulae (\ref{firstOrder}), (\ref{firstOrderCoeff}) constitute the first order approximation to the Krylov basis in this scheme. From this point on we proceed by iteration.  Suppose that the $i^{\mathrm{th}}$ order approximation to the Krylov basis, $|A_n^{(i)}\rangle$ has been computed.  We first find the family of vectors, $|B_n^{(i)}\rangle$, satisfying the orthonormality constraint
\begin{equation}
    \langle A_n^{(i)}| B_m^{(i+1)} 
    \rangle = 0 \ \ \ ; \ \ \ \forall \ n,m   \label{VectorOrth}
\end{equation}
and then the appropriate linear combination of these vectors representing the next order approximation
\begin{eqnarray}
    \left|A_n^{(i+1)}\right\rangle &=& \frac{1}{\sqrt{1 + |N_{i+1,n}|^2 }}\left( \left|A_n^{(i)}\right\rangle + N_{i+1,n} \left|B_{n-\lceil \frac{i}{2}\rceil   }^{(i+1)}\right\rangle    \right)    \nonumber \\
    N_{i+1,n} & = & \frac{ \left\langle B_{n-\lceil \frac{i}{2}\rceil   }^{(i+1)}| H^n|  h, 0, M\right\rangle}{ \left\langle A_n^{(i)} | H^n| h, 0, M  \right\rangle  }   \label{vectorRecurse}
\end{eqnarray}
Note that the vector $|B_n^{(i+1)}\rangle$ is out of step with (i.e. has a different $n$-dependence to) the vector $|A_n^{(i)}\rangle$.  The size of the step, $j$ is determined by finding the the smallest integer for which $\langle h, 0, M| H^n|B_{n-j   }^{(i+1)}\rangle   \neq 0$. 
As with the Lanczos algorithm, this iterative procedure is aimed at obtaining an orthonormal Krylov basis.  However, each step in the iteration process produces not just a single vector, but a family of vectors with increasing precision as an approximation to the true Krylov basis.  The Hamiltonian expressed in this basis is of a tri-diagonal form at each step in the iterative process.  Furthermore, note that the set $\left\{ |A_{n}^{(m)}\rangle \ , \ n=0, 1, 2, \cdots   \right\}$ contains the exact Krylov vector $|K_0\rangle, |K_1\rangle, \cdots |K_{m}\rangle$.  

\subsection{Computation using coherent states}

As was the case in the context of the $su(1,1)$ and Heisenberg Hamiltonians, coherent states provide an efficient computational tool for the implementation of the iterative procedure outlined 
above.  Here again, it is because the vectors are replaced by differential operators acting on holomorphic states, and  overlaps can be computed by acting with these differential operators on scalar functions.} \\ \\ 
As a starting point, note that with judicious usage of the BCH formulas, the time-evolution of the reference state under the Hamiltonian (\ref{JacobiH}) is given by,
\begin{eqnarray}
    & & e^{i t H}|h, 0, M\rangle  \nonumber \\
    & = & \left( \cosh(\sqrt{\zeta} t) - i \frac{i \gamma \sinh(t \sqrt{\zeta})}{2 \sqrt{\zeta}}   \right)^{-2h} \exp\left\{  M \frac{|2 \alpha \beta^* - \beta \gamma|^2 }{4 \zeta^2 \cosh(\sqrt{\zeta} t) - 2 i \zeta^{\frac{3}{2}} \gamma \sinh(\sqrt{\zeta} t)  }  \right\}   \nonumber \\
    & & \exp\left\{ M\left(    - i t \frac{\alpha^* \beta^2 + \alpha (\beta^*)^2 - |\beta|^2 \gamma}{2 \zeta}  +    \frac{\gamma( \alpha^* \beta^2 + \alpha (\beta^*)^2  ) - 4 |\alpha|^2 |\beta|^2}{4|\alpha|^2 \zeta} \right) \right\}    \nonumber \\
    & &   \exp\left\{  M \frac{((4|\alpha|^2 + \gamma^2 )(\alpha^* \beta^2 + \alpha (\beta^*)^2)  - 8 |\alpha|^2 |\beta|^2 \gamma)(4 \sqrt{\zeta} \gamma + i |\alpha|^2 \sinh(2 \sqrt{\zeta} t)  ) }{2 \zeta^{\frac{3}{2}}(16 \zeta^2 - \gamma^4 + |\alpha|^4 \cosh(2 \sqrt{\zeta} t))  }  \right\}     \nonumber \\
        &  & \exp\left\{ \frac{-i(2 \alpha \beta^*(\cosh(\sqrt{\zeta} t) -1) ) - 2 \sqrt{\zeta} \beta \sinh(\sqrt{\zeta} t)}{2 i \zeta \cosh(\sqrt{\zeta} t) + \sqrt{\zeta} \gamma \sinh(\sqrt{\zeta} t)} l_{+}  \right\}  \times  \nonumber \\
    &  & \exp\left\{-\frac{2 \alpha}{\gamma + 2 i \sqrt{\zeta} \coth(\sqrt{\zeta} t)}L_{+} \right\} |h, 0, M\rangle\,,
      \label{tEvolveState}
\end{eqnarray}
with $\zeta \equiv |\alpha|^2 - \frac{\gamma^2}{4}$. This in turn may be combined with (\ref{JacobiCSOL}) to give a bulky though explicit expression for the overlap
\begin{equation}
    F(\bar{z}_1, \bar{z}_2, t) = \langle h, 0, M| e^{\bar{z_2} L_{+}} e^{\bar{z}_1 l_{+}} e^{i t H} |h, 0, M\rangle    \label{OLGenFunc}
\end{equation}
This function serves as a generating functional for the overlaps $\langle A_n^{(i)}| H^n |h,0,M\rangle$ and $\langle B_{n-\lceil \frac{i}{2}\rceil   }^{(i+1)}| H^n |h,0,M\rangle$ needed for (\ref{vectorRecurse}).  In addition, the coherent state overlap (\ref{JacobiCSOL}) and the expectation value of the Hamiltonian
\begin{eqnarray}
& & (\bar{z}_1, \bar{z}_2|H |z_1, z_2) \nonumber \\ & = & e^{M \frac{\bar{z}_1^2 z2  + \bar{z}_2 z_1^2 + 2 \bar{z}_1 z_2}{2(1 - \bar{z}_2 z_2)}}(1 - \bar{z}_2 z_2)^{-2(h+1)} \left( \alpha \left(\frac{M}{2}(\bar{z}_1 + z_1 \bar{z}_2 )^2 + 2 h \bar{z}_2(1 - \bar{z}_2 z_2)   \right) \right.    \nonumber \\
&+& \beta M(\bar{z}_1 + z_1 \bar{z}_2)(1 - \bar{z}_2 z_2) + \gamma\left(  \frac{M}{2}(z_1 + \bar{z}_1 z_2)(\bar{z}_1 + z_1 \bar{z}_2) + h(1 - \bar{z}_2^2 z_2^2 )  \right)   \nonumber \\
&+& \left.  \alpha^* \left(\frac{M}{2}(z_1 + \bar{z}_1 z_2 )^2 + 2 h z_2 (1 - \bar{z}_2 z_2)   \right) +\beta^* M(z_1 + \bar{z}_1 z_2)(1 - \bar{z}_2 z_2)     \right)\,,
\end{eqnarray}
will be important.
In the language of Jacobi coherent states the Krylov basis vectors may be represented by differential operators acting on the state $|z_1, z_2)$.  This is analogous to the analysis performed for the $SU(1,1)$ and Heisenberg groups in subsection \ref{KExampleSection}.  Here the "0$^{\mathrm{th}}$" order Krylov vectors correspond to the differential operator
\begin{eqnarray}
    \hat{A}_n^{(0)} & = & \frac{e^{-i a}}{\sqrt{4h|\alpha|^2 + 2 n M |\beta|^2} \sqrt{n! (2h+1)_{(n-1)}}}\left( \alpha \partial_{z_2}^n + \beta n \partial_{z_2}^{n-1} \partial_{z_1}     
    \right)\,,
\end{eqnarray}
with $|A_n^{(0)}\rangle = \hat{A}_n^{(0)} |z_1, z_2)|_{z_1 \rightarrow 0, z_2 \rightarrow 0}$. Written in the coherent state language, the algorithm now proceeds as follows:  Given $\hat{A}_n^{(i)}$ such that
\begin{equation}
     \left. (\hat{A}_m^{(i)})^\dag \hat{A}_n^{(i)} (\bar{z}_1, \bar{z}_2|z_1, z_2) \right|_{z_1, \bar{z}_1 \rightarrow 0, z_2, \bar{z}_2 \rightarrow 0} = \delta_{m,n}\,,
\end{equation}
we need to find the family of differential operators $\hat{B}_n^{(i+1)}$ such that
\begin{equation}
         \left. (\hat{B}_m^{(i+1)})^\dag \hat{A}_n^{(i)} (\bar{z}_1, \bar{z}_2|z_1, z_2) \right|_{z_1, \bar{z}_1 \rightarrow 0, z_2, \bar{z}_2 \rightarrow 0} = 0\,,\label{DiffOrth}
\end{equation}
for all $n,m$. The next order approximation is then given by 
\begin{eqnarray}
    \hat{A}_n^{(i+1)} &=& \frac{1}{\sqrt{1 + |N_{i+1,n}|^2 }}\left( \hat{A}_n^{(i)} + N_{i+1,n} \hat{B}_{n-\lceil \frac{i}{2}\rceil   }^{(i+1)}\rangle    \right)\,,    \nonumber \\
    N_{i+1,n} & = & \left( \left.\hat{B}^{i+1}_{n-\lceil \frac{i}{2}\rceil} \partial_t^n F(\bar{z}_1, \bar{z}_2, t) \right|_{\bar{z}_1 \rightarrow 0, \bar{z}_2 \rightarrow 0, t\rightarrow 0}  \right)   \left( \left.\hat{A}^{i}_{n} \partial_t^n F(\bar{z}_1, \bar{z}_2, t) \right|_{\bar{z}_1 \rightarrow 0, \bar{z}_2 \rightarrow 0, t\rightarrow 0}  \right)^{-1}\,.   \label{DiffRecurse}   \nonumber
\end{eqnarray}
When terminated at order $j$ the Lanczos coefficients may thenbe extracted as
\begin{eqnarray}
b^{(j)}_n & = & (\hat{A}^{(j)}_{n})^\dag (\hat{A}^{(j)}_{n-1}) \left. (\bar{z}_1, \bar{z}_2 | H | z_1, z_2)   \right|_{z_1,\bar{z}_1 \rightarrow 0, z_2,\bar{z}_2 \rightarrow 0}   \nonumber \\
a^{(j)}_n & = & (\hat{A}^{(j)}_{n})^\dag (\hat{A}^{(j)}_{n}) \left. (\bar{z}_1, \bar{z}_2 | H | z_1, z_2)   \right|_{z_1,\bar{z}_1 \rightarrow 0, z_2,\bar{z}_2 \rightarrow 0} 
\end{eqnarray}

\section{Jacobi Lanczos coefficients and complexity}

We are now in a position to compute the approximate Lanczos coefficients for the  Hamiltonian \eqref{JacobiH}.  In the next few sections we will compute these at 0$^{\mathrm{th}}$, 1$^{\mathrm{st}}$ and 2$^{\mathrm{nd}}$ order for some choices of parameters and compare these to the exact Lanczos coefficients which we determine numerically. 
Following this we will be able to compute the corresponding approximate Krylov complexities.  A nice feature is that, since the terms appearing in the sum are known for all values of $n$, we have good control over the accuracy of the results.   
\subsection{Lanczos Coefficients}

We now proceed to apply the algorithm outlined above up to second order, setting $h=\frac{1}{4}$ in line with (\ref{hMCondition}). For ease of notation, we set
\begin{equation}
w_n = \sqrt{|\alpha|^2 + 2 M n |\beta|^2 }.
\end{equation}
We find the families of differential operators
\begin{eqnarray}
 \hat{B}_n^{(1)} & = & \frac{\alpha^*} {\alpha \sqrt{M} \sqrt{16 h^2 \alpha^2 + 8 n h M \beta^2} \sqrt{(n-1)! (2h+1)_{(n-1)}}  }\left(  2 M \beta^* \partial_{z_2}^n - 4 \alpha^* h \partial_{z_2}^{n-1} \partial_{z_1} \right)  \nonumber \\
 A_n^{(1)} & = &  \frac{1}{\sqrt{1 + |N_{2,n}|^2 }}\left( \hat{A}_n^{(0)}  + N_{2,n} \hat{B}^{(1)}_{n-1}   \right)   \nonumber \\
 B_n^{(2)} & = & \frac{1}{\sqrt{1 + |N_{2,n}|^2 }}\left( N_{2,n} \hat{A}_n^{(0)}  -  \hat{B}^{(1)}_{n-1}   \right)   \nonumber \\
 A_n^{(2)} & = &  \frac{1}{\sqrt{1 + |N_{3,n}|^2 }}\left( \hat{A}_n^{(1)}  + N_{3,n} \hat{B}^{(2)}_{n-1}   \right)  
\end{eqnarray}
with coefficients
\begin{equation}
N_{2,k} \nonumber \\
 =  \frac{\sqrt{M k(k-1) }(4(2k-1)M\beta^3 - 6 \alpha^2 \beta^* + 3 \alpha \beta \gamma ) ) }{6 \alpha \sqrt{2k-1}  w_k w_{k-1} } 
\end{equation}
and   
\begin{eqnarray}
N_{3,k} & = & \frac{\sqrt{M \ k (k-1)(k-2)}}{180 \sqrt{2}\alpha^3 \beta \sqrt{   (1 + |N_{2,k-1}|^2 ) (1 + |N_{2,k-1}|^2 )  }\sqrt{(2k-3)(2k-1)} w_k w_{k-1}^2 w_{k-2}     } \times \nonumber \\
& & \left[   (5 |\alpha|^2 + w_{k-1}^2)(4 M \beta^3(2k-1) - 6 \alpha^2 \beta^* + 3 \alpha \beta \gamma )(4 M \beta^3(2k-3) - 6 \alpha^2 \beta^* + 3 \alpha \beta \gamma )    \right.     \nonumber \\
& & \left.  - 36 \alpha (2 \alpha \beta^* - \beta \gamma)( 2 M \beta^3(k-1) + 3 \alpha^2 \beta^* + \alpha \beta \gamma  ) w_{k-1}^2    \right]    \nonumber
\end{eqnarray}
These expressions may be readily obtained using a simple Mathematica code implementing the algorithm of the previous subsection.  The Lanczos coefficients may subsequently be obtained as 
\begin{eqnarray}
b_{n}^{(2)} & = & \frac{\sqrt{n(n-\frac{1}{2})} w_n \alpha}{\sqrt{(1 + |N_{2,n-1}|^2)(1 + |N_{2,n}|^2)(1 + |N_{3,n-1}|^2)(1 + |N_{3,n}|^2)} w_{n-1} }   \nonumber \\
&+& \frac{  \sqrt{(2n-3)(n-2)} w_{n-1}w_{n-2} N_{2,n-1}N_{2,n}^* \alpha  +   \sqrt{M(n-1)}(4(n-1) M \beta^3 - 2 \alpha^2 \beta^* + \alpha \beta \gamma )N_{2,n}^*   }{\sqrt{2}\sqrt{(1 + |N_{2,n-1}|^2)(1 + |N_{2,n}|^2)(1 + |N_{3,n-1}|^2)(1 + |N_{3,n}|^2)} w_{n-1}^2 }   \nonumber \\
&+& \frac{N_{3,n-1} N_{3,n}^* \left(   \sqrt{(2n-5)(n-3)}  w_{n-3}w_{n-2} \alpha  +   \sqrt{(2n-3)(n-1)} w_{n-1} w_{n-2} N_{2,n-1} N_{2,n-2}^* \alpha     \right)}{\sqrt{2}\sqrt{(1 + |N_{2,n-2}|^2)(1 + |N_{2,n-1}|^2)(1 + |N_{3,n-1}|^2)(1 + |N_{3,n}|^2)}w_{n-2}^2}    \nonumber \\
&-&  \frac{N_{3,n-1} N_{3,n}^* \sqrt{M(n-1)} \left( 4(n-2) M \beta^3 - 2 \alpha^2 \beta^*  + \alpha \beta \gamma  \right) N_{2,n-2}^*   }{\sqrt{2}\sqrt{(1 + |N_{2,n-2}|^2)(1 + |N_{2,n-1}|^2)(1 + |N_{3,n-1}|^2)(1 + |N_{3,n}|^2)}w_{n-2}}   \nonumber \\
&+& \frac{N_{3,n}^*( -2\sqrt{(n-2)(2n-3)(n-1)} w_{n-1} w_{n-2} M^\frac{3}{2} |\beta|^2 (\beta - \beta^* N_{2,n-1}^2)  + \frac{1}{2} \gamma w_{n-1}^2 w_{n-2}^2 N_{2 n-1}  )}{(1 + |N_{2,n-1}|^2)\sqrt{(1 + |N_{3,n-1}|^2)(1 + |N_{3,n}|^2)}w_{n-1}^2 w_{n-2}^2}    \nonumber \\
&+&  \frac{N_{3,n}^*N_{2,n-1}( 2 M ( (2n-2)w_{n-2}^2 - |\alpha|^2  )(\alpha^* \beta^2 + \alpha (\beta^*)^2 )   +  \gamma |\alpha|^2 w_{n-\frac{3}{2}}^2  )}{(1 + |N_{2,n-1}|^2)\sqrt{(1 + |N_{3,n-1}|^2)(1 + |N_{3,n}|^2)}w_{n-1}^2 w_{n-2}^2}   \label{LanczosCoeff}
\end{eqnarray}
The expression is organised so that the first line represents our 0$^{\mathrm{th}}$ order approximation, the second line the 1$^{\mathrm{st}}$ order approximation and the rest the 2$^{\mathrm{nd}}$ order approximation.  The 0$^{\mathrm{th}}$ order may be extracted from the above by setting $N_{2,n}, N_{3,n} \rightarrow 0$ and the 1$^{\mathrm{st}}$ order by setting $N_{3,n} \rightarrow 0$.  These expressions are bulky, reflecting the involved combinatorics determining the Krylov basis.  Nevertheless, these are explicit expressions that we may test numerically.  We may also compare these with an explicit analytical computation of the first few orders of the algorithm, as described in appendix \ref{LanczosAppendix}.  \\ \\
\begin{figure}[h]
\begin{minipage}{0.49 \textwidth}
    \centering
    \subfloat[$|\beta| = \frac{1}{4}$]{\includegraphics[width=0.9\textwidth]{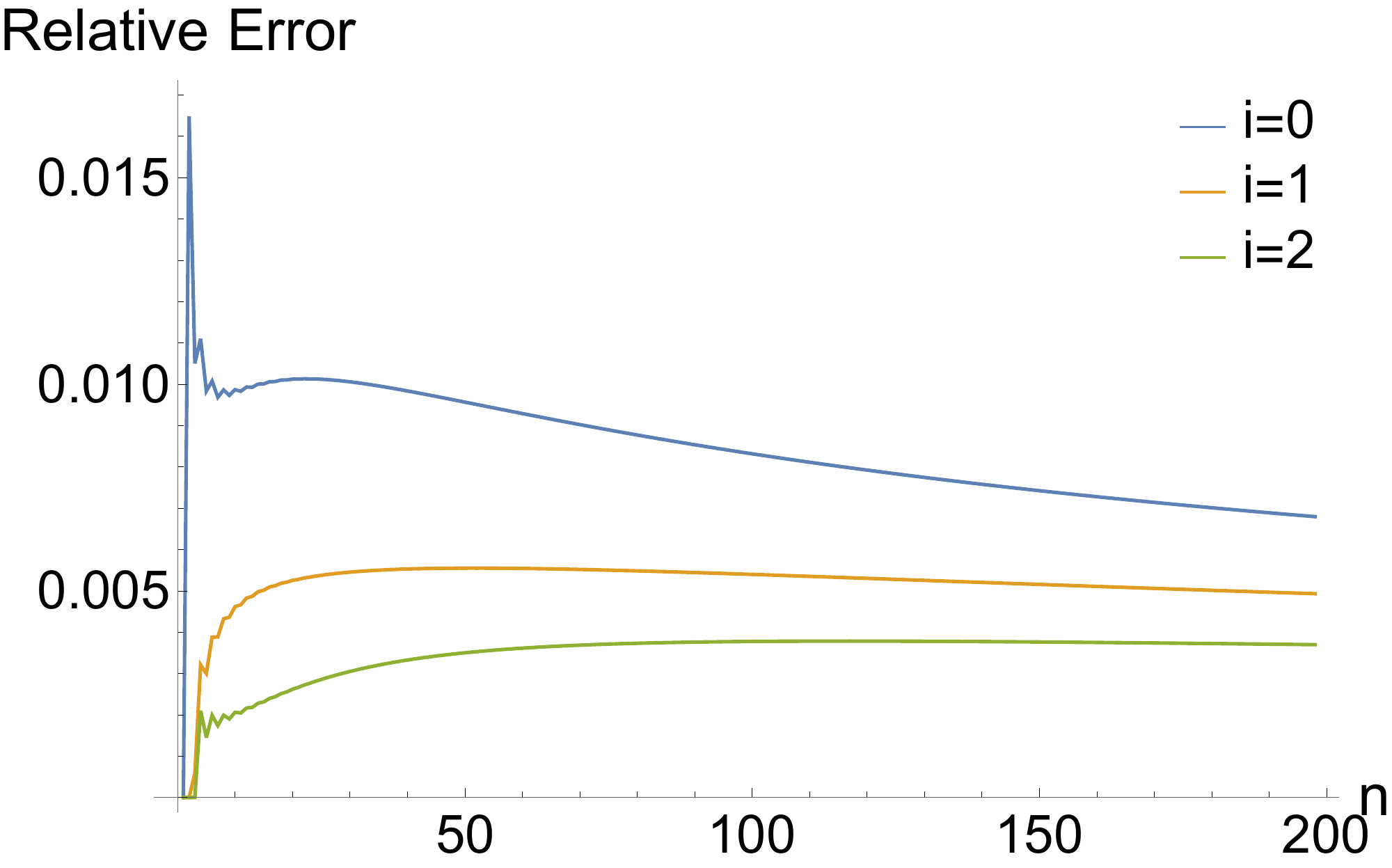} }
    \vfill
     \subfloat[$|\beta| = \frac{1}{2}$]{\includegraphics[width=0.9\textwidth]{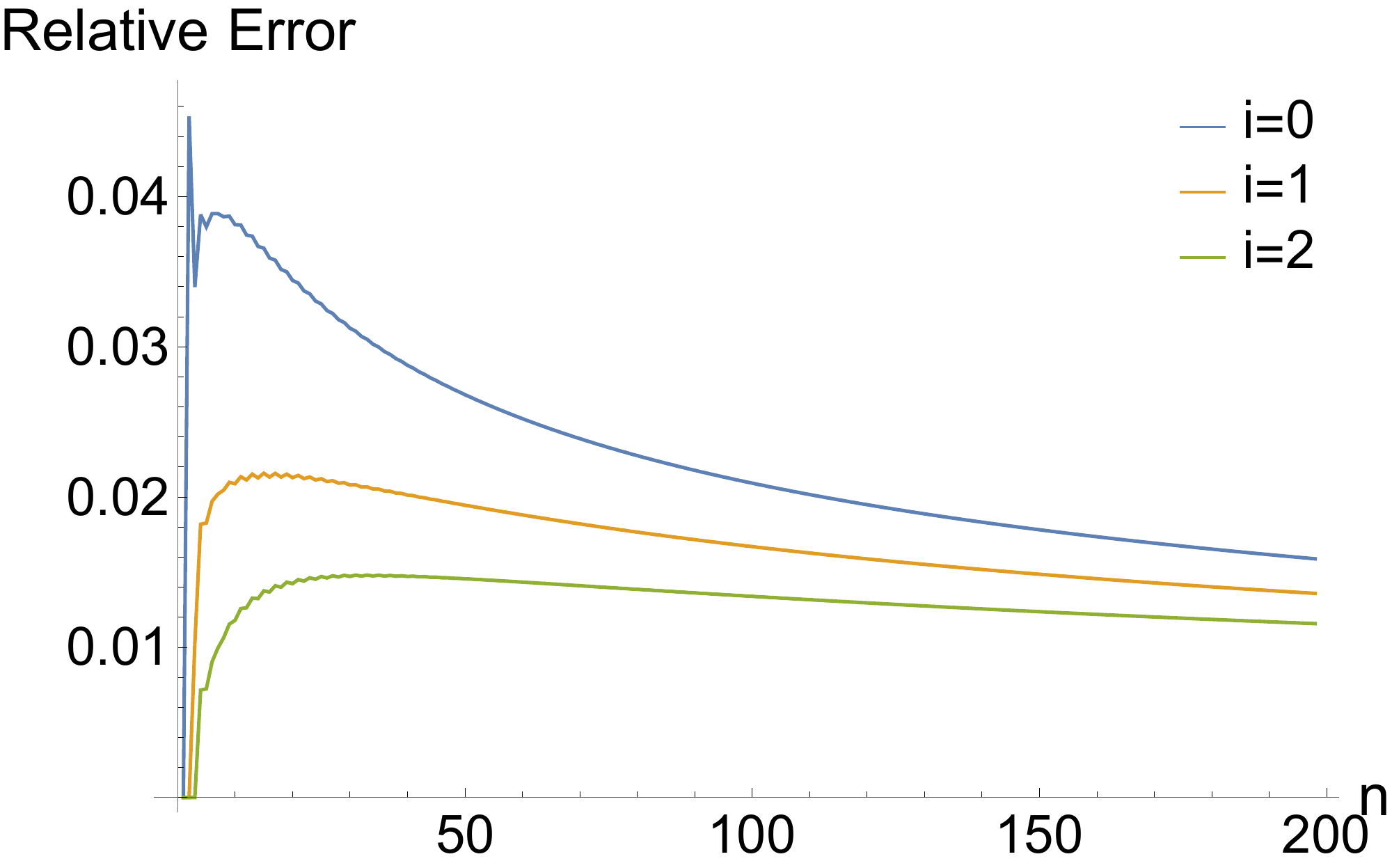} }
\end{minipage}
\begin{minipage}{0.49 \textwidth}
    \centering
    \subfloat[$|\beta| = 1$]{\includegraphics[width=0.9\textwidth]{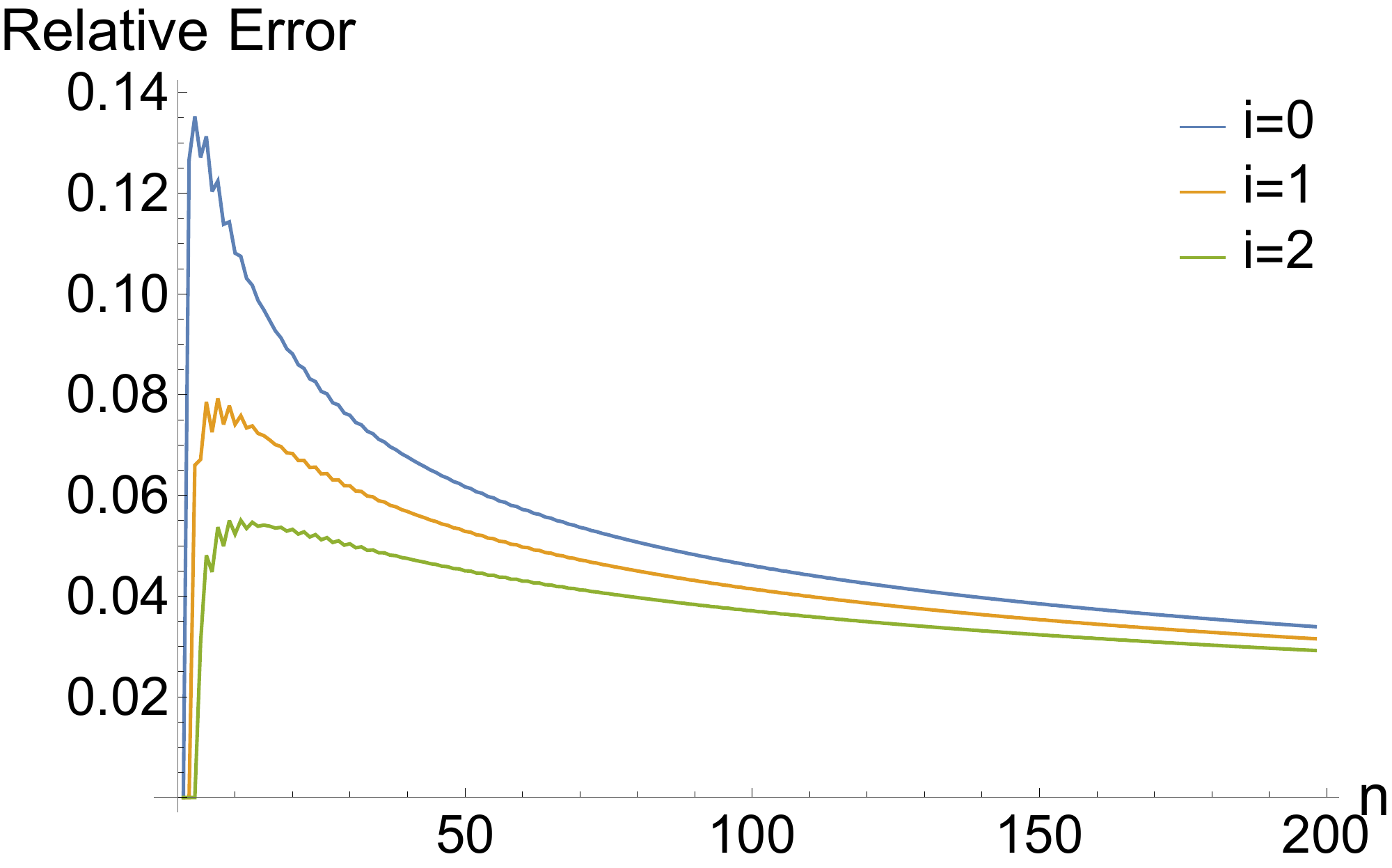} }
    \vfill
     \subfloat[$|\beta| = 2$]{\includegraphics[width=0.9\textwidth]{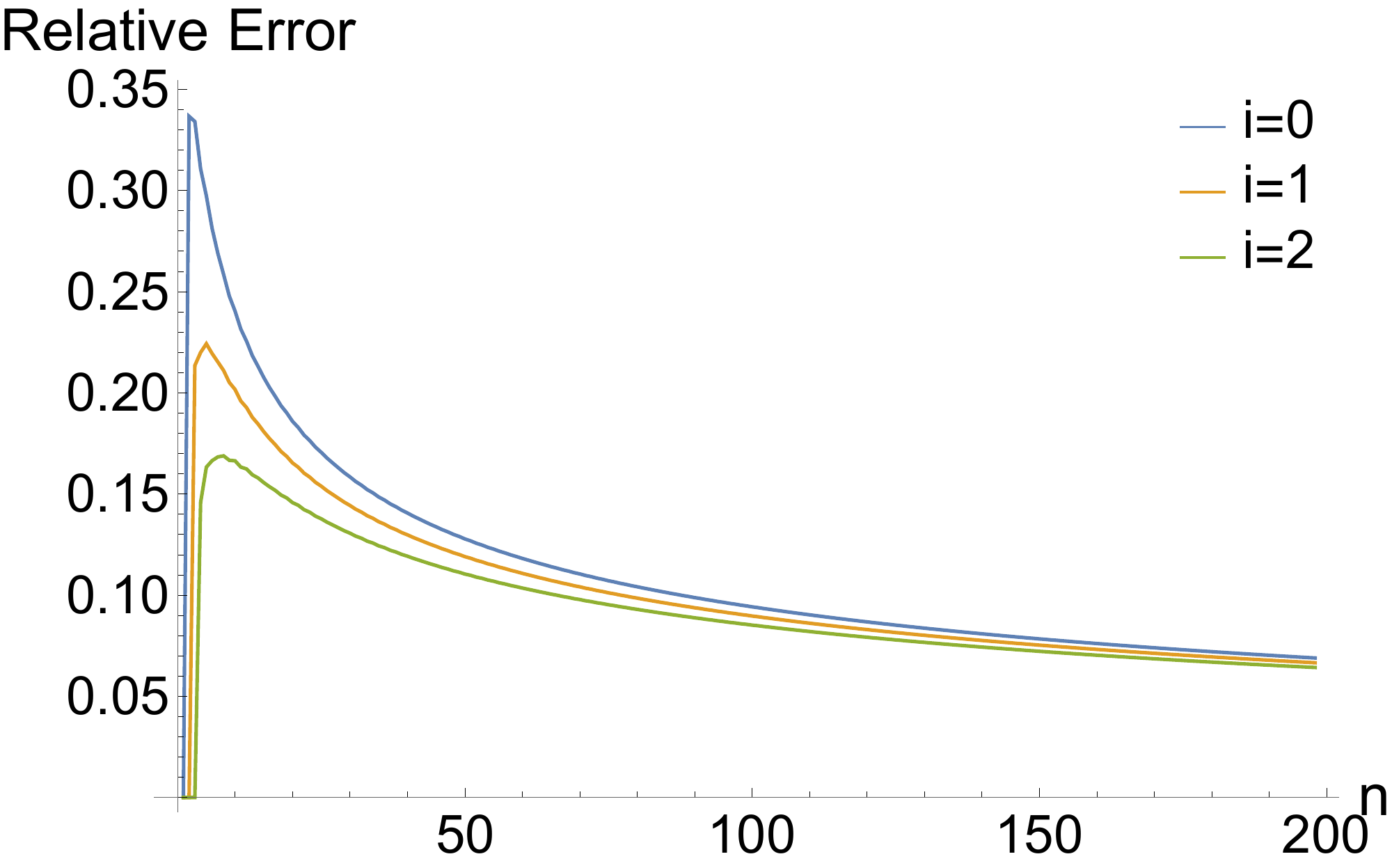} }
\end{minipage}
    \caption{The relative error for the Lanczos coeffcients $\left|   \frac{b_n - b_{n}^{i}}{b_n}   \right|$ for $i=0$ (blue), $i=1$ (orange) and $i=2$ (green) with $|\alpha| =1, \gamma = 0, \frac{\alpha}{\alpha^*} = \frac{\beta}{\beta^*} = e^{2 i \frac{\pi}{3}}$. }
    \label{LanczosCoeff1}
\end{figure}
\begin{figure}[h]
\begin{minipage}{0.49 \textwidth}
    \centering
    \subfloat[$\gamma = \frac{1}{4}$]{\includegraphics[width=0.9\textwidth]{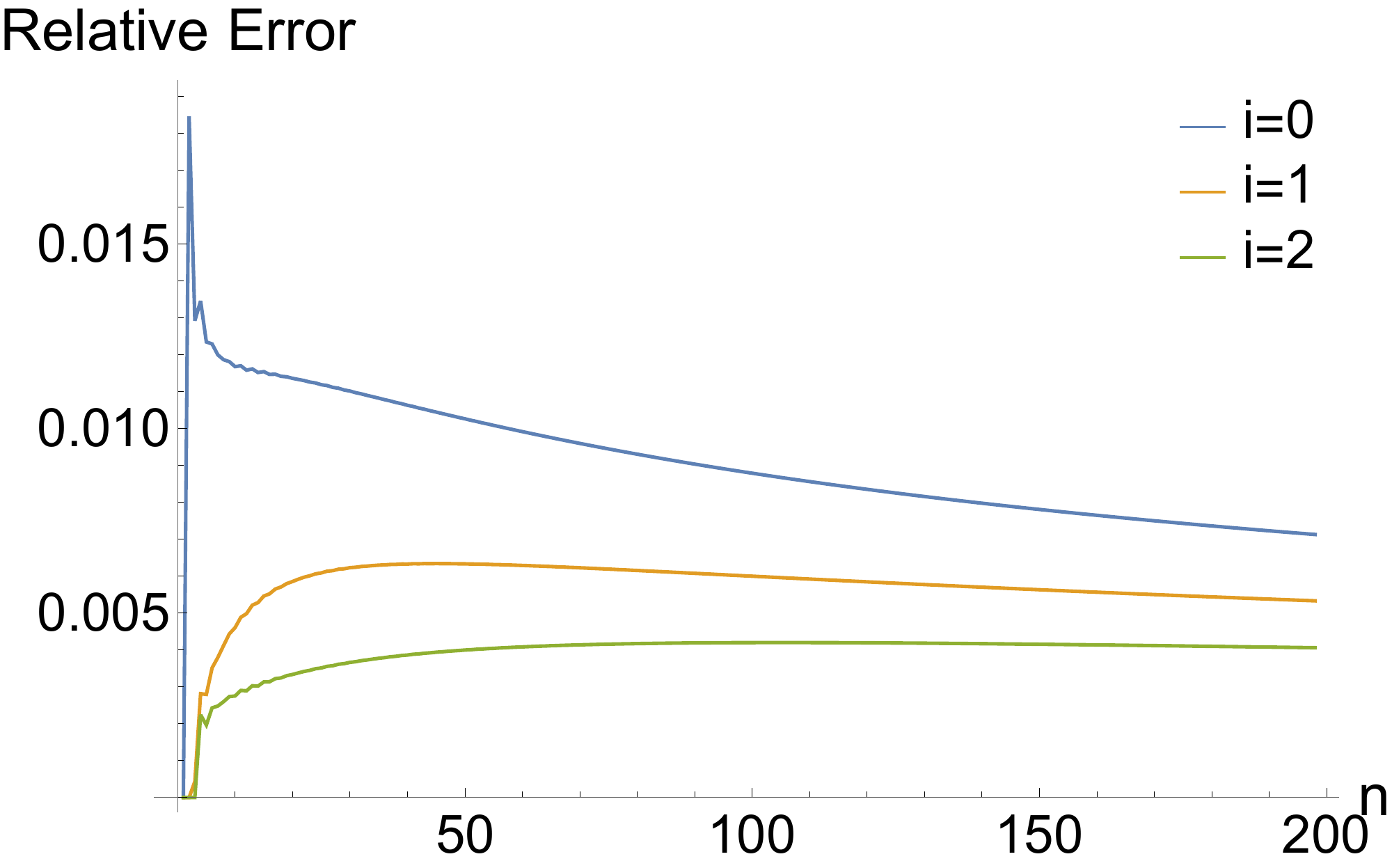} }
    \vfill
     \subfloat[$\gamma = \frac{1}{2}$]{\includegraphics[width=0.9\textwidth]{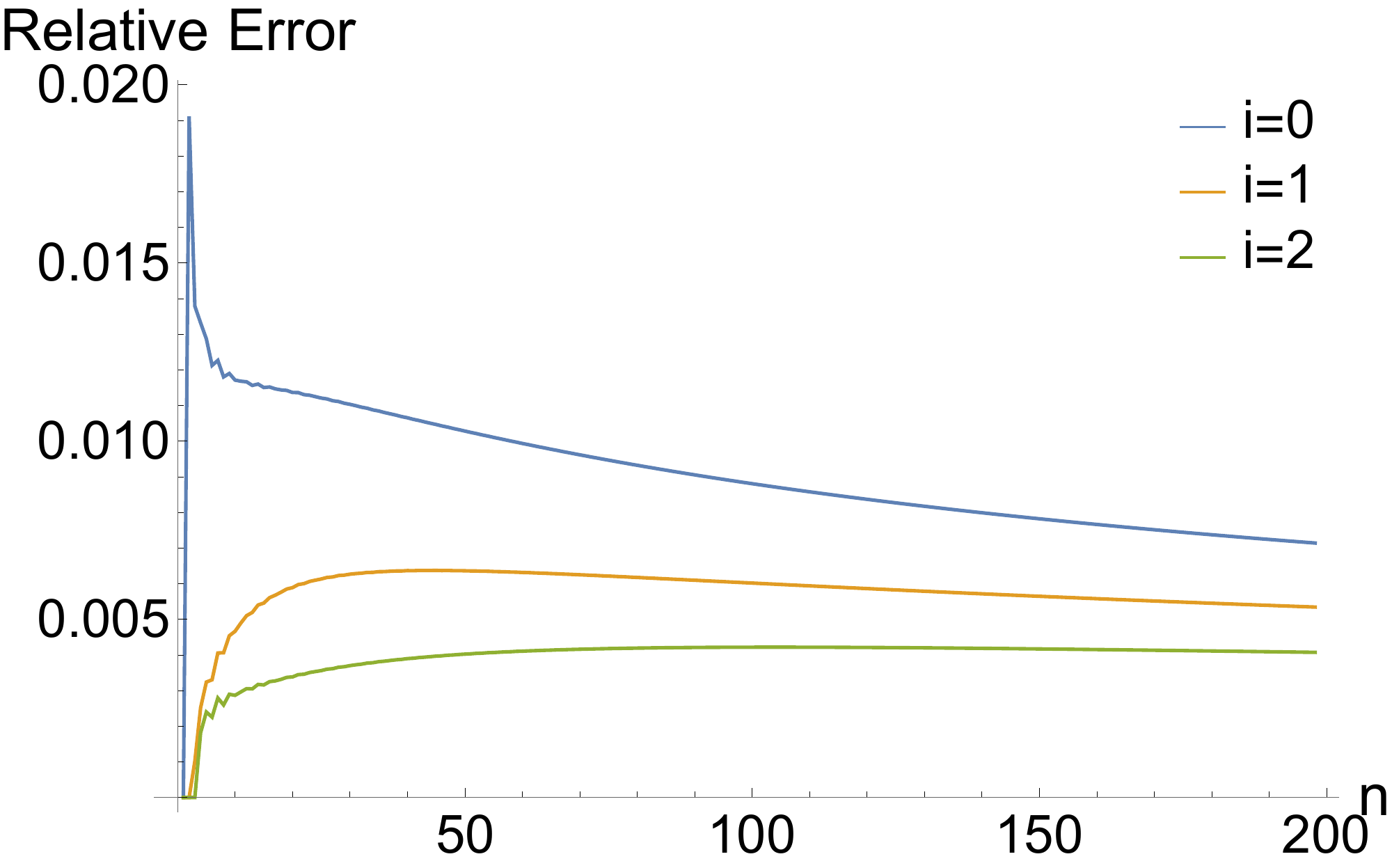} }
\end{minipage}
\begin{minipage}{0.49 \textwidth}
    \centering
    \subfloat[$\gamma = 1$]{\includegraphics[width=0.9\textwidth]{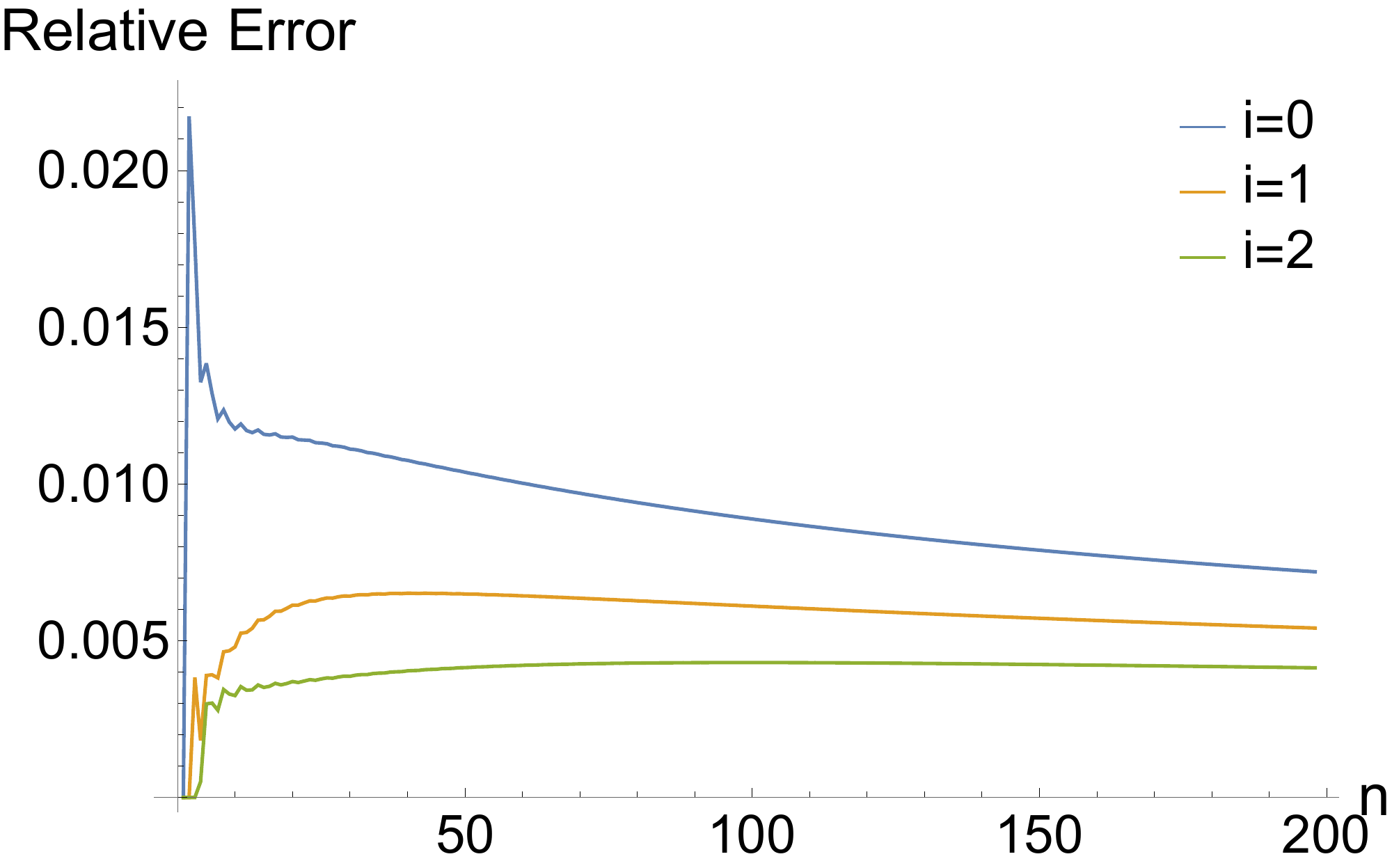} }
    \vfill
     \subfloat[$\gamma = 2$]{\includegraphics[width=0.9\textwidth]{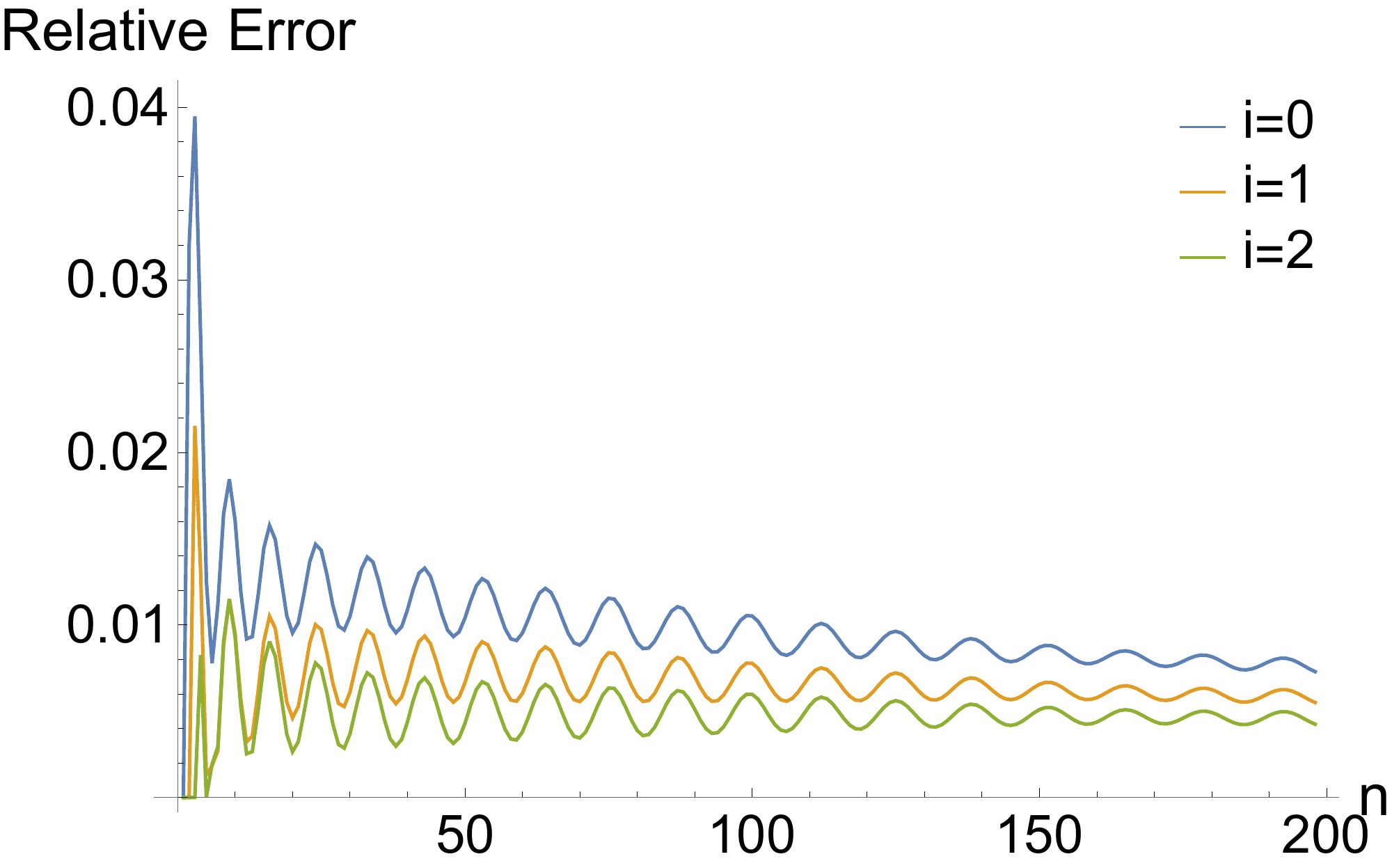} }
\end{minipage}
    \caption{The relative error for the Lanczos coeffcients $\left|   \frac{b_n - b_{n}^{i}}{b_n}   \right|$ for $i=0$ (blue), $i=1$ (orange) and $i=2$ (green) with $|\alpha| =1, \beta = \frac{1}{4}, \frac{\alpha}{\alpha^*} = e^{i \pi},  \frac{\beta}{\beta^*} = 1$. }
    \label{LanczosCoeff2}
\end{figure}
In Figs. (\ref{LanczosCoeff1}) and (\ref{LanczosCoeff2}) we have plotted the relative error of terminating the algorithm at 0$^{\mathrm{th}}$, 1$^{\mathrm{st}}$ and 2$^{\mathrm{nd}}$ order.  The Lanczos coefficients are computed numerically and compared with (\ref{LanczosCoeff}).  We note that, even for values $|\beta| \approx |\alpha|$ the leading order approximation provides an error of less than $10$ percent with greater accuracy at smaller values of $|\beta|$.   At subsequent orders the approximation increases in accuracy, especially for smaller values of $n$.  As advertised, the algorithm perfectly reproduces $b_{j+1}$ at order $j$.   

\subsection{K-complexity}

Given a Krylov basis $\{|K_{n}\rangle\}$, we define the Krylov complexity operator as,
\begin{equation}
    \hat{K} = \sum_n n |K_n \rangle \langle K_n |.
\end{equation}
The associated K-complexity of a particular target state is then given by the expectation value of this operator in this state.  In our case, we will use as the target state the time-evolved reference state (\ref{tEvolveState})
which (up to an overall normalisation) is a Jacobi coherent state with values
\begin{eqnarray}
    z_1(t) & = & \frac{(2 \alpha \beta^* - \beta \gamma)(\cosh(\sqrt{\zeta} t) -1) + 2 i \beta \sqrt{\zeta} \sinh(\sqrt{\zeta} t) }{2 \zeta \cosh(\sqrt{\zeta } t) - i \gamma \sqrt{\zeta} \sinh(\sqrt{\zeta} t)}\,,   \nonumber \\
    z_2(t) & = & -\frac{2 \alpha}{\gamma + 2 i \sqrt{\zeta} \coth(\sqrt{\zeta} t)}\,,   \label{z1z2Vals}   \\
    \zeta & \equiv & |\alpha|^2 - \frac{\gamma^2}{4}\,.     \nonumber
\end{eqnarray}
In terms of the differential operators we write
\begin{equation}
    K^{(i)} = \sum_{n} n \frac{1}{(\bar{z}_1(t), \bar{z}_2(t) |  z_1(t), z_2(t)  )}\left. ( A_{n}^{(i)} )^\dag ( A_{n}^{(i)} ) \left( \bar{z}_1, \bar{z}_2 | z_1(t), z_2(t) \right) \left( \bar{z}_1(t), \bar{z}_2(t) | z_1, z_2 \right)    \right|_{z_1,z_2\rightarrow 0}   \label{Ki}
\end{equation}
which is the approximate spread complexity at "i$^{\mathrm{th}}$" order.  A useful formula to compute the above is 
\begin{eqnarray}
 \left. \partial_{\bar{z}_2}^n \partial_{\bar{z}_1}^{2m} (\bar{z}_1, \bar{z}_2| z_1, z_2) \right|_{\bar{z}_1 \rightarrow 0, \bar{z}_2 \rightarrow 0}
   &=& \sum_{k=0}^{n+m} 2^{m -k} \left( \begin{array}{c} n+m \\ k \end{array}\right) \frac{\Gamma(n + m + \frac{1}{2})}{\Gamma(k + \frac{1}{2})} z_1^{2k} z_2^{n+m-k}  \nonumber \\
   \left. \partial_{\bar{z}_2}^n \partial_{\bar{z}_1}^{2m+1} (\bar{z}_1, \bar{z}_2| z_1, z_2) \right|_{\bar{z}_1 \rightarrow 0, \bar{z}_2 \rightarrow 0} 
   &=& \sum_{k=0}^{n+m} 2^{m -k} \left( \begin{array}{c} n+m \\ k \end{array}\right) \frac{\Gamma(n + m + \frac{3}{2})}{\Gamma(k + \frac{3}{2})} z_1^{2k+1} z_2^{n+m-k} \nonumber
\end{eqnarray}
which may be written formally as a hypergeometric function.   \\ \\ 
 Before proceeding further, it is worthwhile to reflect on the physical expectations for Krylov complexity.  The Hamiltonians we consider are similar to harmonic oscillator and inverted harmonic oscillator studied previously in the complexity literature \cite{Ali:2019zcj,Bhattacharyya:2020art, Haque:2021kdm} (see also the $su(1,1)$ example in \cite{Caputa:2021sib, Haque:2021hyw}).  Indeed, when using the creation / annihilation operator representation, these systems are related by a shift of the creation / annihilation operators and Bogoliubov transformation.  The frequency of the oscillator is given by $\omega = \sqrt{\frac{\gamma^2}{4} - |\alpha|^2}$.  \\ \\
When $\omega$ is real we expect complexity to be a periodic function in time.  Note that this feature is already guaranteed by the parametrisation of the time-evolved state as a Jacobi coherent state (\ref{z1z2Vals}).  For real $\omega$ the hyperbolic trigonometric functions become ordinary trigonometric functions.  In contrast, when $\omega$ is imaginary we expect complexity similar to that of the inverted oscillator which should grow exponentially.  This may again be expected by the form of the parametrisation (\ref{z1z2Vals}) which saturates at late times and, furthermore, the large $n$ behavior of the Lanczos coefficients (\ref{LanczosCoeff}) is $b_n^{2} = \alpha n + o(1)$.  Put together we  expect exponential growth of Krylov complexity at late time.   \\ \\
\begin{figure}
\begin{minipage}{0.32 \textwidth}
    \centering
    \subfloat[$a = 0$]{\includegraphics[width=0.9\textwidth]{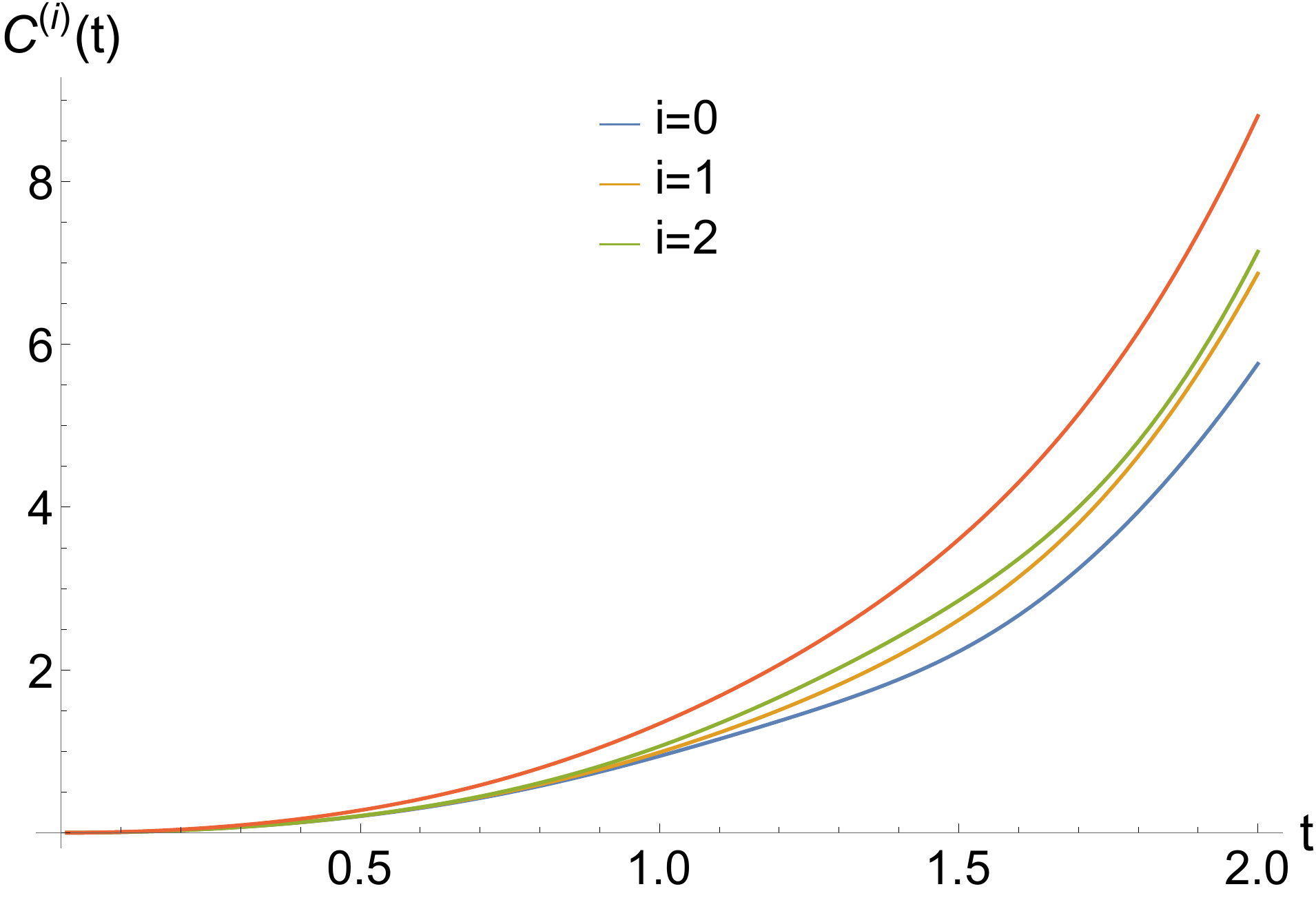} }
\end{minipage}
\begin{minipage}{0.32 \textwidth}
    \centering
    \subfloat[$a = \frac{2 \pi}{3}$]{\includegraphics[width=0.9\textwidth]{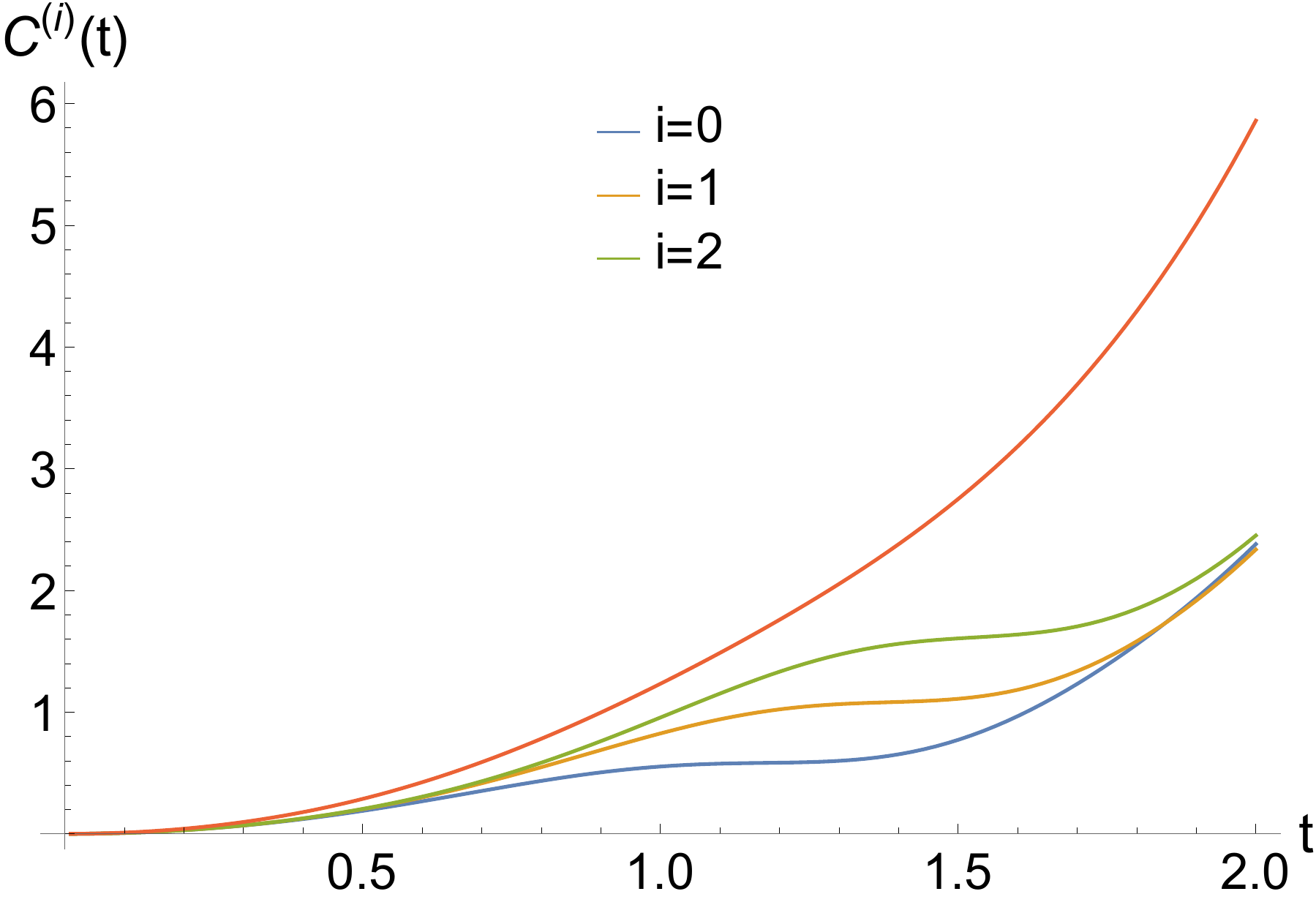} }
\end{minipage}
\begin{minipage}{0.32 \textwidth}
    \centering
    \subfloat[$a = \frac{4\pi}{3}$]{\includegraphics[width=0.9\textwidth]{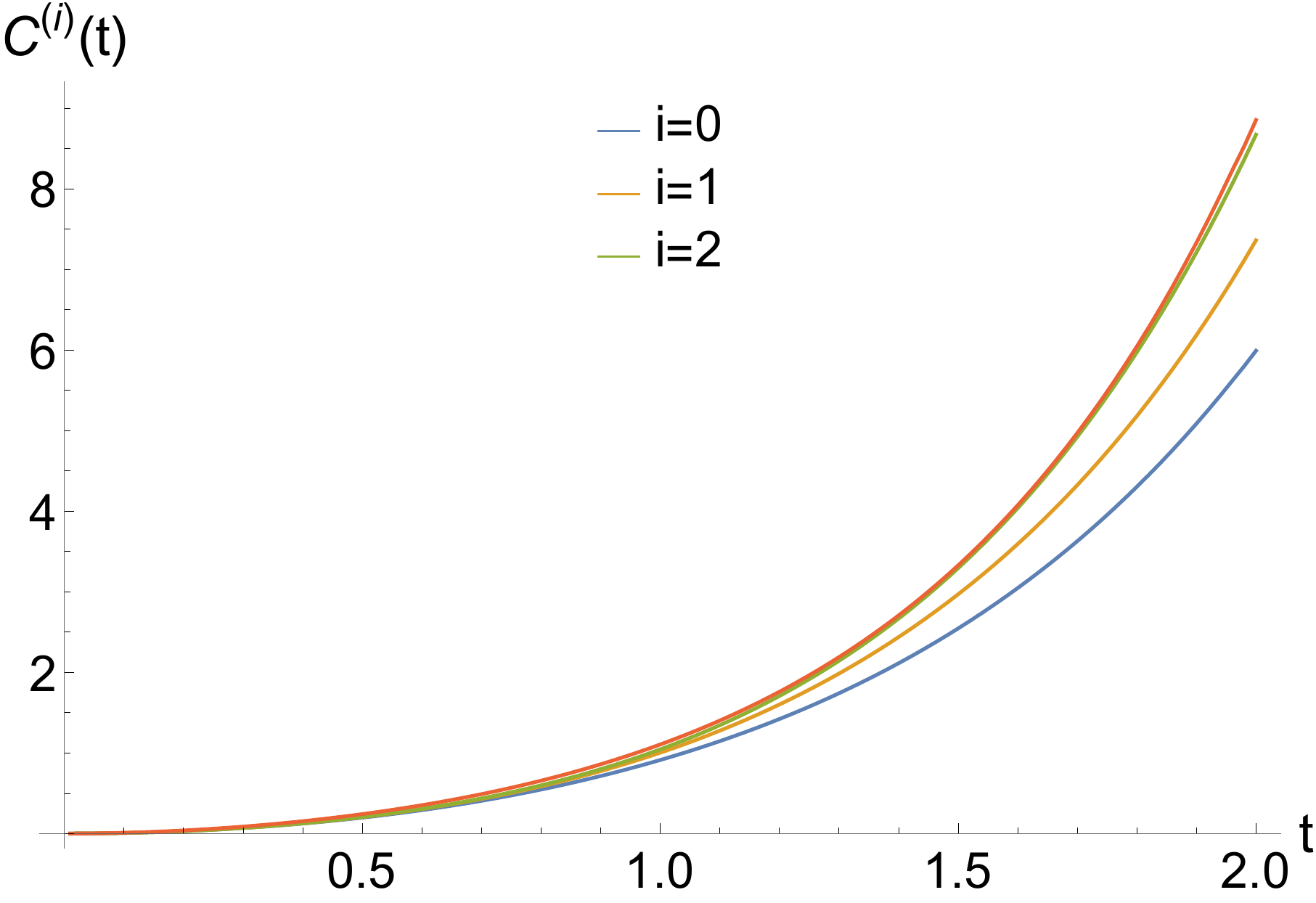} }
\end{minipage}
    \caption{The approximate spread complexity, $K^{(i)}$ of the time-evolved reference state at leading, subleading and subsubleading order in the scheme.  We have taken $\alpha = 1$, $\beta = \frac{1}{2}$, $\gamma = 1$ and $b = \frac{\pi}{2}$.  The red line is a plot of the expectation value (\ref{CostFunctilde}) for comparison.}
    \label{CompFig1}
\end{figure}

\begin{figure}
\begin{minipage}{0.32 \textwidth}
    \centering
    \subfloat[$\beta = \frac{1}{4}$]{\includegraphics[width=0.9\textwidth]{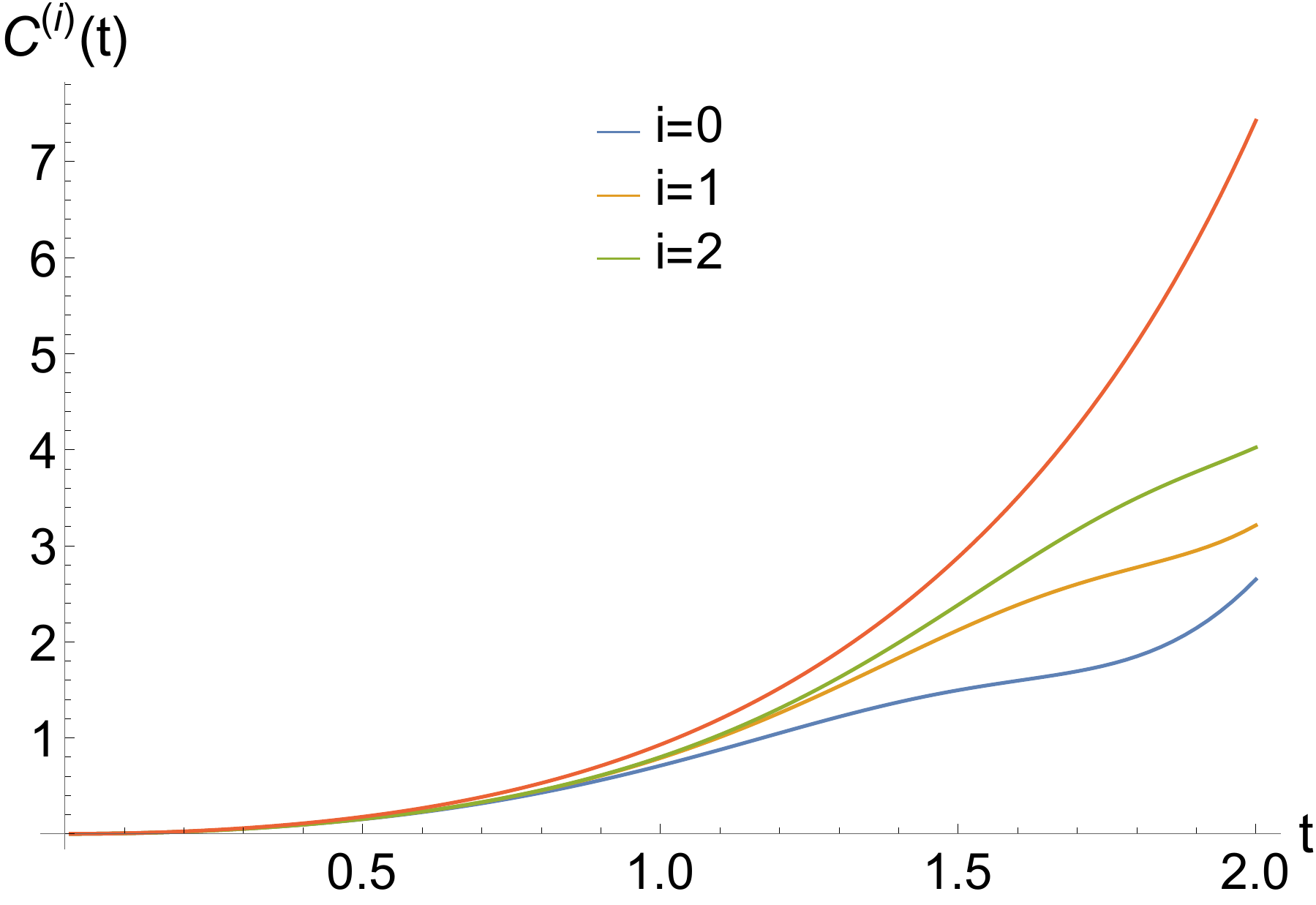} }
\end{minipage}
\begin{minipage}{0.32 \textwidth}
    \centering
    \subfloat[$\beta = \frac{1}{2}$]{\includegraphics[width=0.9\textwidth]{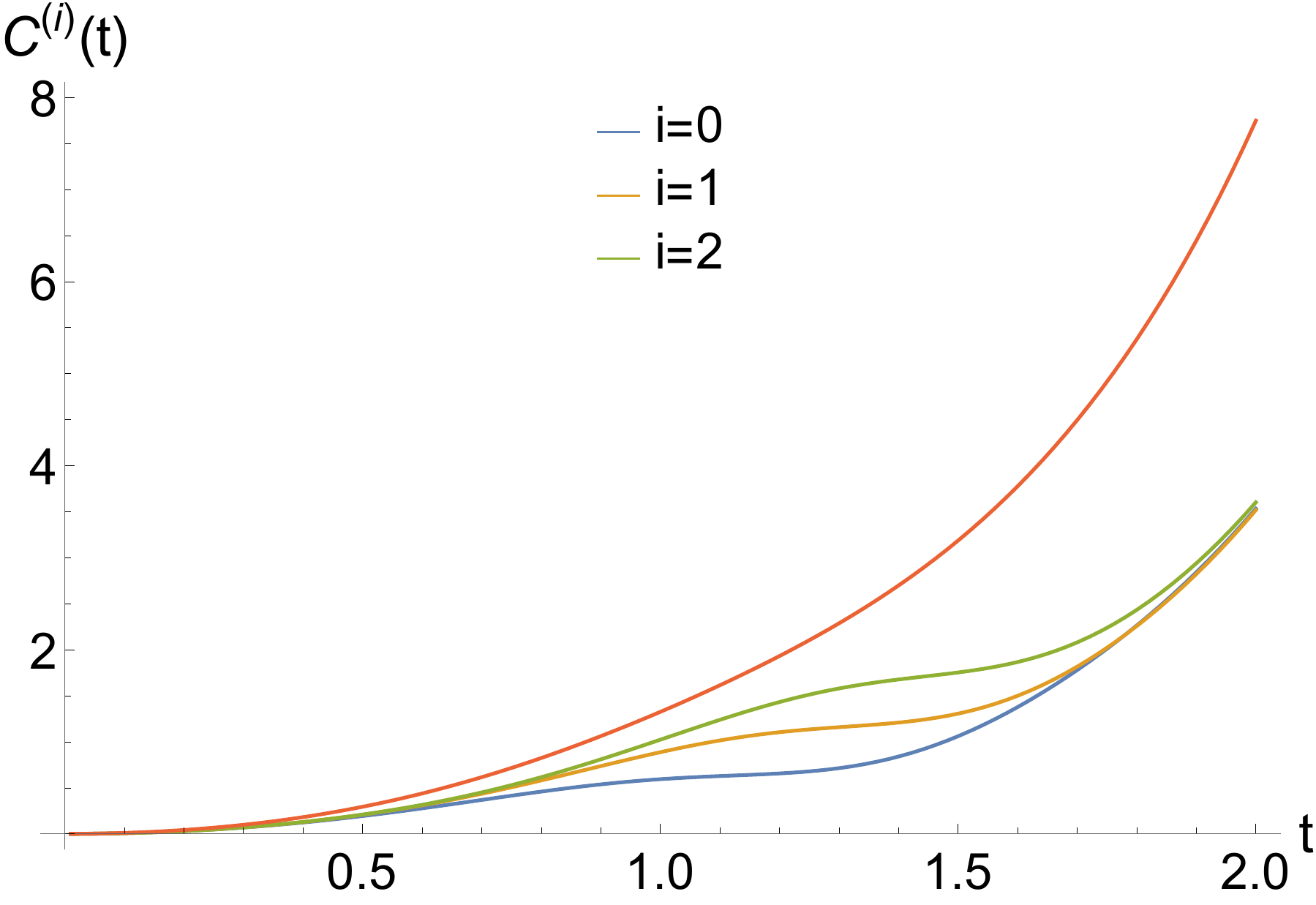} }
\end{minipage}
\begin{minipage}{0.32 \textwidth}
    \centering
    \subfloat[$\beta = 1$]{\includegraphics[width=0.9\textwidth]{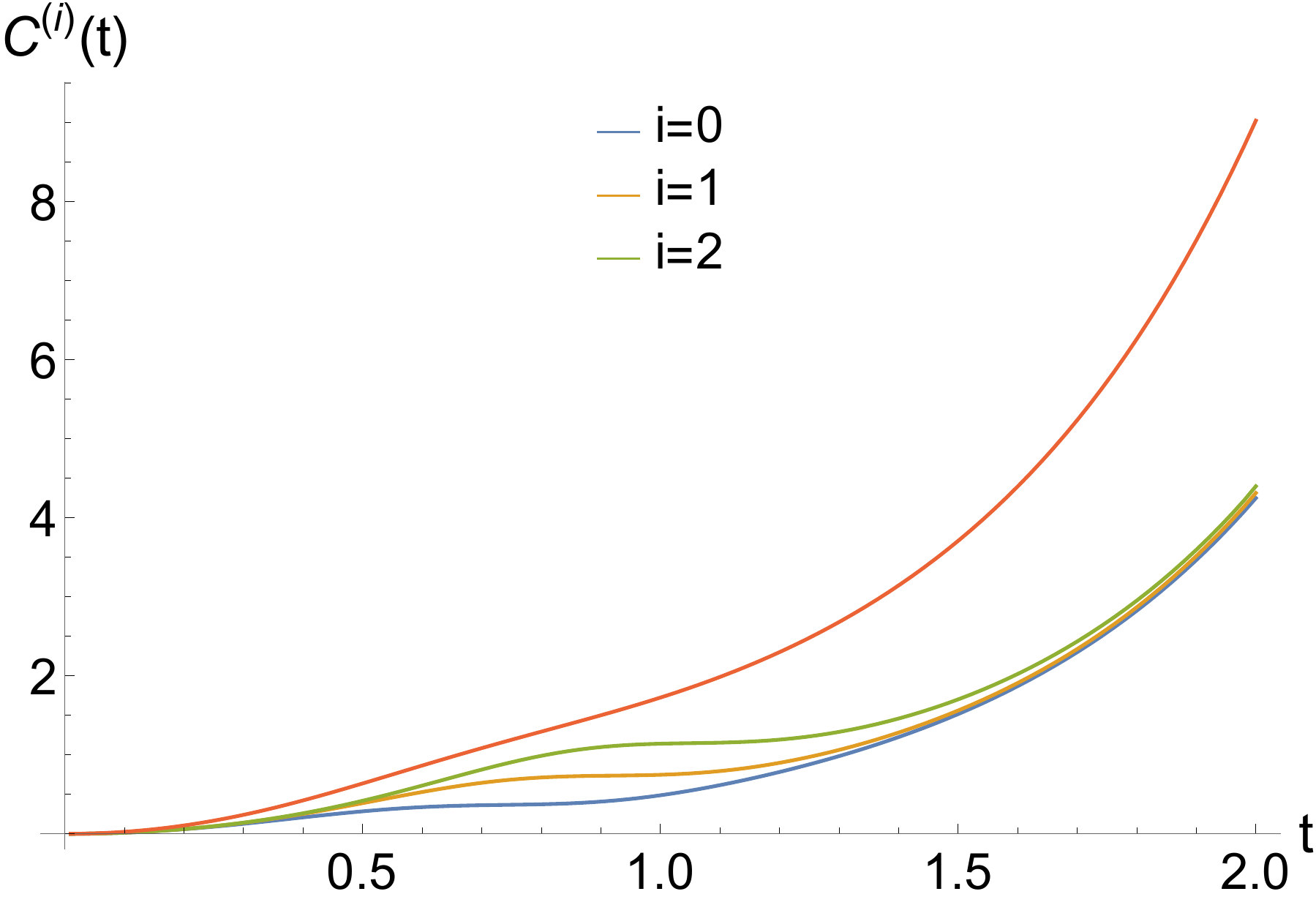} }
\end{minipage}
    \caption{The approximate spread complexity, $K^{(i)}$ of the time-evolved reference state at leading, subleading and subsubleading order in the scheme.  We have taken $\alpha = 1$, $\gamma = 0$ and $a=b = \frac{\pi}{2}$.  The red line is a plot of the expectation value (\ref{CostFunctilde}) for comparison.}
    \label{CompFig2}
\end{figure}
\begin{figure}
\begin{minipage}{0.32 \textwidth}
    \centering
    \subfloat[$a = 0$]{\includegraphics[width=0.9\textwidth]{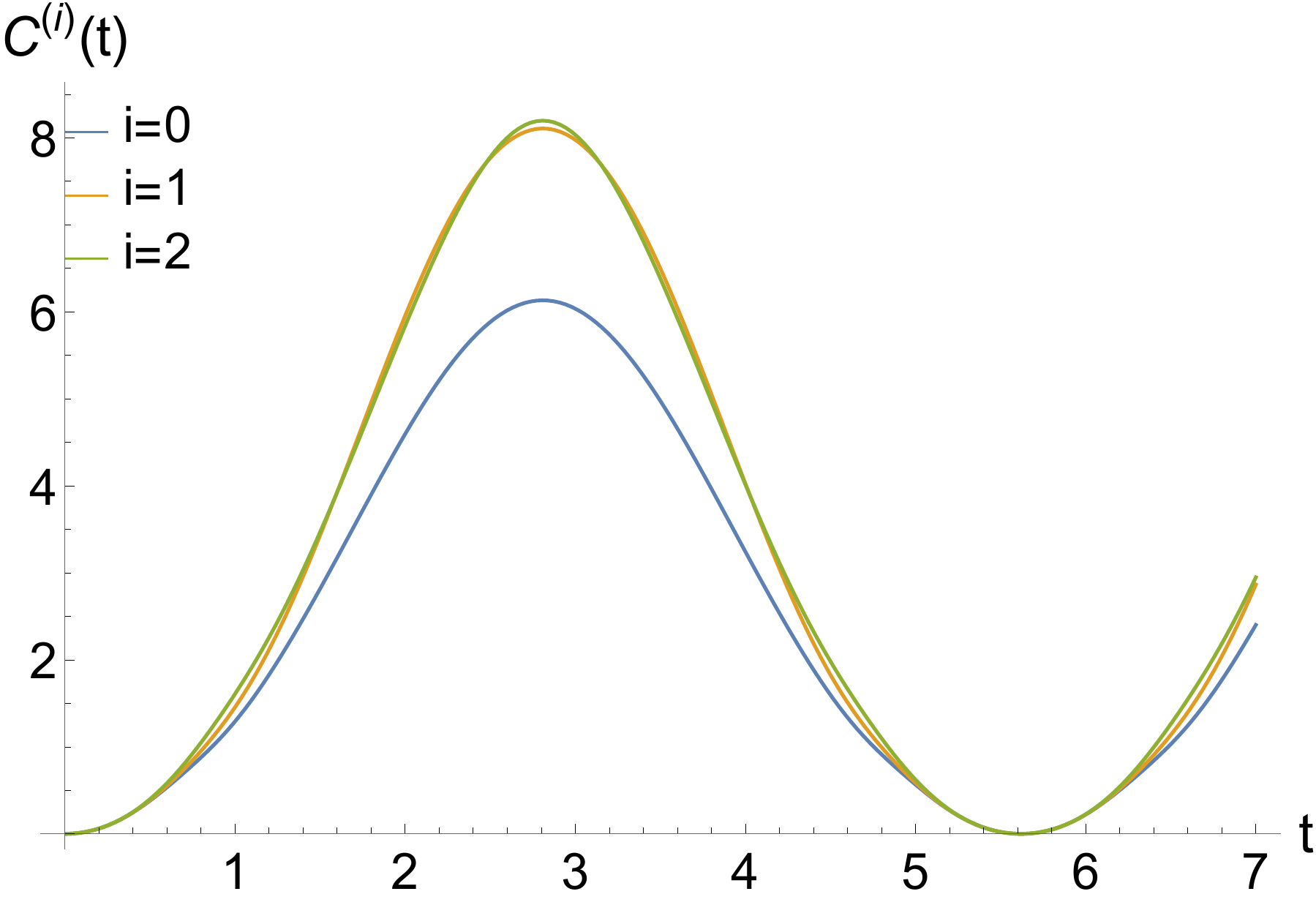} }
\end{minipage}
\begin{minipage}{0.32 \textwidth}
    \centering
    \subfloat[$a = \frac{\pi}{3}$]{\includegraphics[width=0.9\textwidth]{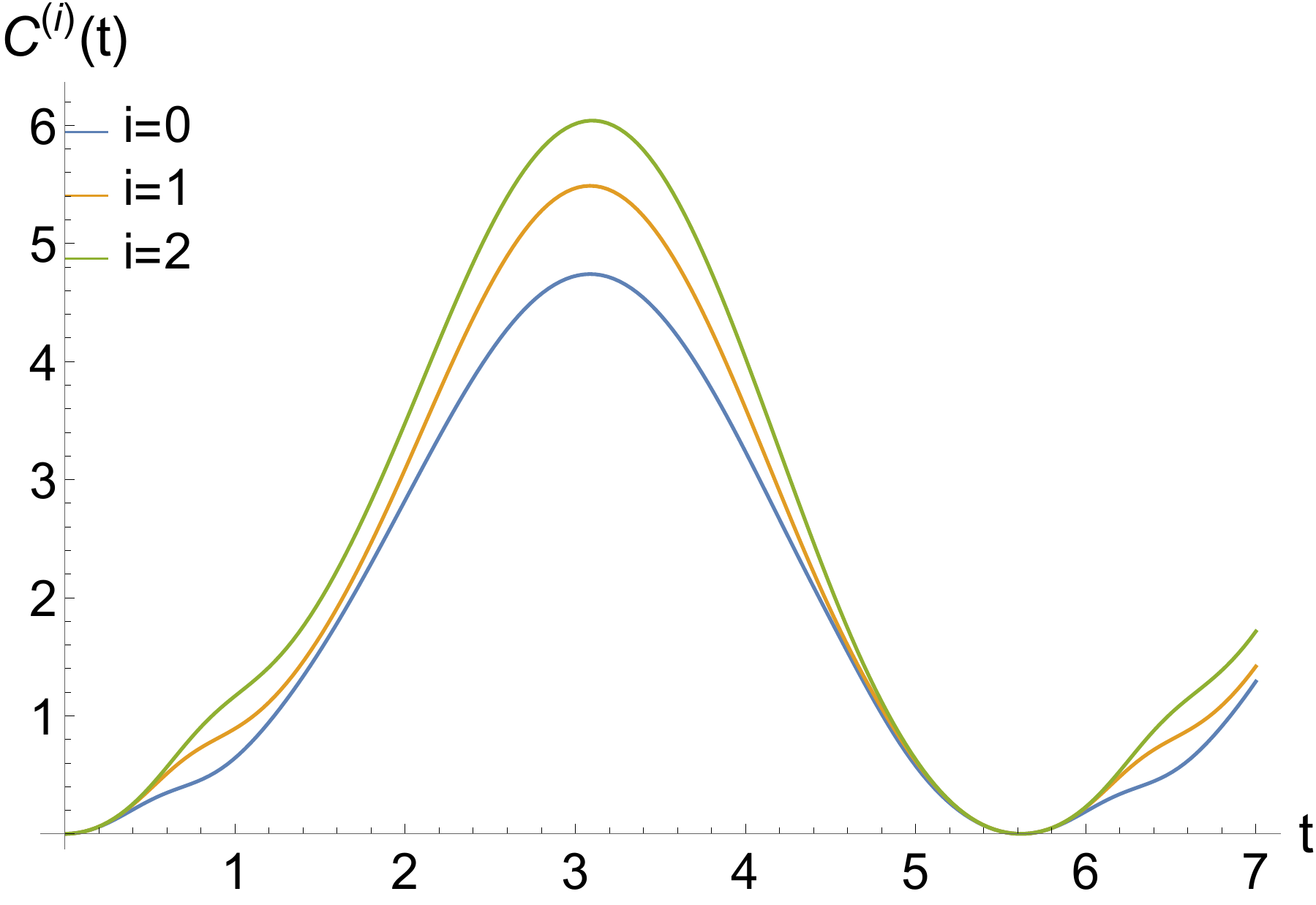} }
\end{minipage}
\begin{minipage}{0.32 \textwidth}
    \centering
    \subfloat[$a = \frac{2 \pi}{3}$]{\includegraphics[width=0.9\textwidth]{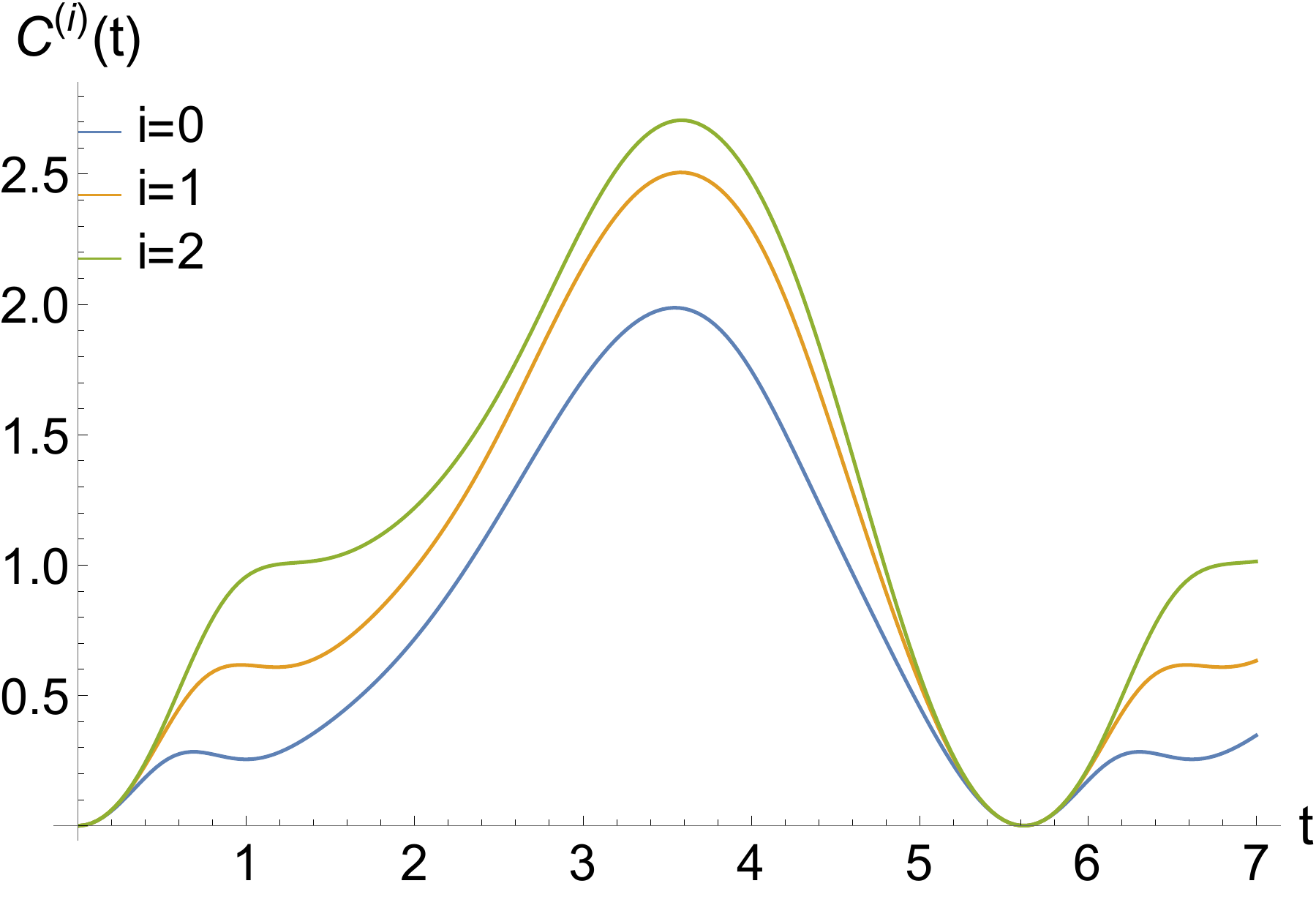} }
\end{minipage}
    \caption{The approximate spread complexity, $K^{(i)}$ of the time-evolved reference state at leading, subleading and subsubleading order in the scheme.  We have taken $|\alpha| = |\beta| = 1$, $\gamma = 3$ and $b = \frac{\pi}{2}$. A little more than one period of the periodic approximate spread complexity is shown.}
    \label{CompFig3}
\end{figure}
\noindent
In Figs. (\ref{CompFig1}), (\ref{CompFig2}) we have plotted the spread complexity for a range of parameters corresponding to an imaginary frequency while those in Fig. (\ref{CompFig3}) correspond to real frequency.  For comparison we have plotted the expectation value
\begin{equation}
\tilde{C} = \sum_{k}  
  k |\langle h,k, M
 |z_1(t), z_2(t)\rangle|^2+    (k+2) |\langle h, k+\frac{1}{2}, M
 |z_1(t), z_2(t)\rangle|^2  \label{CostFunctilde}
\end{equation}
which is always greater than (\ref{Ki}) for $i=0,1,2$ for all parameter choices that we have numerically checked.
 The physical relevance of this quantity needs some unpacking.  First, note that it is slightly larger than half the value for the cost function \begin{equation}
C_B = \sum_{n} n \langle B_n |z_1(t), z_2(t)\rangle
\end{equation}
with $|B_n\rangle = |h,\frac{n}{2}, M\rangle$.  This is the basis one would use with only access to the Heisenberg generators. However,  since the time evolution involves the $su(1,1)$ generators (which ladder up faster than the Heisenberg generators) it is easy to understand why the complexity is less than the cost function, $C_B$.
Indeed, the expectation value (\ref{CostFunctilde}) may be interpreted as a cost function with the basis generated by the $su(1,1)$ generators from a "complexity $0$ state" $|h, 0, M\rangle$ and a "complexity $2$ state" $|h, \frac{1}{2}, M\rangle$.  This value of $2$ is allocated since both $H|h, 0, M\rangle$ and $H^2|h, 0, M\rangle$ are needed to produce both relevant linear combinations of $|h, \frac{1}{2}, M\rangle$ and $|h, 1, M\rangle$ present in the problem\footnote{If the state $|h, \frac{1}{2}, M\rangle$ is assigned a "complexity" of $1$ the cost function may be smaller than $K^{(i)}$ at early times.}.  The complexity we obtain is less than $\tilde{C}$.  One way to think about this is that the Heisenberg generators provide additional ways to synthesise the target state. 
 For a choice of parameters corresponding to imaginary frequencies we observe that the complexity is an increasing function of time with exponential growth at later times. 
 If, instead, the parameters are chosen to correspond to a real frequency, the complexity is a periodic function in time.  Both of these results are in line with our physical expectations.  
\\ \\
Another noteworthy feature we would like to highlight is the appearance of an inflection point in $K^{(i)}$ at intermediate times and for some choices of parameters, see Figs. (\ref{CompFig1}), (\ref{CompFig2}).    Though we do not yet fully understand its appearance, intuitively,  it may be understood as the result of the fact that a particular target state is obtainable in various ways using the ladder operators $L_{+}$ and $l_{+}$.  As we follow the time-evolved state, its decomposition into the various Krylov basis components will thus involve many powers of $z_1(t)$ and $z_2(t)$, which grow differently as functions of time.  As such, it is possible for the gradient of complexity to change, depending on the precise coefficients appearing in front of various powers of $z_1(t)$ and $z_2(t)$.  
 A similar feature may be noted for a family of geodesics in the context of Nielsen complextiy \cite{Nielsen_2006, Nielsen1, Nielsen3}, where the tangent vector to evolution on the space of unitaries may be an arbitrary combination of $l_{+}$ and $L_{+}$.  The geodesic length may increase differently along these different directions (for a simple example of this see appendix B).  \\ \\
What may be surprising is that the appearance of the inflection point appears to be sensitive to the phases $a,b$, even for the leading order $K^{(0)}$.  This is despite the Lanczos coefficient at this order (\ref{LanczosCoeff}) not being sensitive to the phase.  Here it is important to stress that, though the operator $K^{(i)}$ is an approximate K-complexity operator, the states $|z_1(t), z_2(t)\rangle$ are the exact time-evolved states, and the dependence on the phases enters through it.  Indeed, we have noted that, in order for the inflection point to appear, a necessary condition is that $|z_1(t)|$ is comparable to $|z_2(t)|$ at intermediate times.  

\subsection{Generalisations}

We would like to emphasise that the recursive method, in vector form in (\ref{VectorOrth}), (\ref{vectorRecurse}) and in differential form in (\ref{DiffOrth}), (\ref{DiffRecurse}), is a general algorithm to generate the Krylov basis.   Provided that a natural guess for a ``$0^{\mathrm th}$ order'' basis exists, one may iterate to generate a sequence of bases that converge on the Krylov basis.  We have applied this in the context of Jacobi coherent states and the resulting expressions capture the involved combinatorics associated with the Gram-Schmidt procedure.  The Jacobi group provides three simplifications that would not be available in a more general setting.  We would like to comment on the application of our proposed algorithm in the absence of these simplifications.  \\ \\   Firstly, for the Jacobi group we were able to obtain an explicit realisation of the time-evolved state in terms of Jacobi coherent states (\ref{tEvolveState}).  This was used to obtain a generating function (\ref{OLGenFunc}) for the overlaps of $H^n|\psi_r\rangle$ with the $|A_n^{(i)} \rangle$ vectors in the approximation scheme.  For generic Hamiltonians this would not be possible though the action of $H^n$ may always be represented in terms of differential operators acting on coherent states.  These overlaps can  still be obtained, even without the convenient generating functional.   \\ \\
Secondly, for a Hamiltonian from the Jacobi algebra the dimension of the span of $\left\{ H^n |\phi_0\rangle \right\}$ is exactly half that of $\left\{ |h, n, M\rangle, |h, n-\frac{1}{2}, M\rangle \right\}$.   Because of this the vectors $|B_n^{(i)}\rangle$ could be obtained with relative ease at each step in the recursive algorithm.  In general there could be several families of vectors orthogonal to $|A_n{(i)}\rangle$ at each step.  In this case one would have to find appropriate linear combinations of these 
vectors, similar to what was performed for the $N_{i+1,n}$ coefficients.  \\ \\
Finally, our Hamiltonian was chosen as an element of some finite-dimensional algebra.  General Hamiltonians will not be of this type.  However, it may be possible to select a reference state which transforms trivially under many terms present in the Hamiltonian.  In such cases, a natural choice for the ``$0^{\mathrm{th}}$'' order basis should be possible.

\section{Discussion}

Spread complexity - a measure of the difficulty of preparing a desired target quantum state from a simple reference state - has emerged as a useful diagnostic for various quantum phenomena, from chaos to phase transitions.  It naturally involves the spectrum of the system under consideration and requires only the reference state, target state and system Hamiltonian to be specified.  Finding the Krylov basis associated with a particular choice of Hamiltonian and reference state is one of the central tasks in computing the complexity.  For simple symmetry groups the Krylov basis can readily be identified in terms of the ladder operators.  In general systems, the Lanczos algorithm provides an iterative way to implement the required Gram-Schmidt orthogonalisation procedure.  \\ \\
In this article we have proposed a modification of the Lanczos algorithm which we anticipate will be well-suited to a variety of problems.  Starting from a basis assembled from from the ladder operators present in the problem, the basis may be iteratively refined until ultimately converging on the Krylov basis.  A key advantage of this modification is that an approximate basis for the Krylov subspace is available at each step in the iteration.  For an infinite-dimensional Krylov subspace this is especially useful to approximate the infinite sums involved in computing the spread complexity in a stable way.  \\ \\
As an illustration, we have applied the algorithm to Jacobi coherent states, where some degree of analytical control is available.  Analytic expressions may be obtained for the approximate Lanczos coefficients at each step of the algorithm and we have demonstrated that the approximation is already accurate at subsubleading order for a range of parameters.  With the aid of analytic expressions, the approximate Krylov may be computed with a simple numerical algortihm.  The behavior we obtain is in accord with physical expectations. 
Notably, we have found that the K-complexity for the semi-direct product group is {\it less} than that of its constitutent subgroups. At first sight, this seems counter intuitive. One way to think about this is by comparison to the equivalent result in circuit complexity\footnote{In Appendix B, we illustrate some features of the circuit complexity for the Jacobi group.} where, in enlarging the symmetry group we are effectively adding more gate sets thereby creating new pathways in the space of unitaries with potentially shorter geodesics between the initial and target states. In this sense then, quantum systems with this more generic quadratic Hamiltonian are less complex than systems evolved by the constituent subsystem Hamiltonians. We anticipate that this result might have interesting practical applications for quantum computations. We will report on these findings in detail in a future work. \\ \\
\noindent
We anticipate that this algorithm may be applied to Hamiltonians drawn from more general symmetry groups and may even be applicable to problems where perturbative treatments are unavoidable.  One example that would be of great interest is the $d$-dimensional conformal group $SO(d,2)$. Results for the Nielsen complexity of conformal primaries are known \cite{Chagnet:2021uvi, Koch:2021tvp, Rabambi:2022jwu}, and it would be interesting to compare these results with Krylov complexity.  A key question in this context is whether the spread-complexity of a time-evolved state under an $SO(d,2)$ Hamiltonian saturates and, if so, how dependent is this property on the choice of reference state.   \\ \\ 
It would be useful to develop a better physical intuition for the information contained in the spread complexity of a quantum state.  For example, one might like to know whether (and under what circumstances) spread complexity is sensitive to phase transition or integrable-to-chaotic transitions. 
  These have been investigated for some models, e.g. \cite{ Jian:2020qpp, Kim:2021okd, Rabinovici:2021qqt, Rabinovici:2022beu, Caputa:2022eye, Caputa:2022yju} (see also \cite{Bhattacharjee:2022vlt}), but the list may be extended substantially.  The tools we have developed in this paper are taking a step towards performing spread complexity computations for these more general cases.   
\\ \\
On a more technical level, the qualitative results for the Lanczos coefficients and spread complexity of the Jacobi group coherent states appear in good agreement with the (numerically obtained) exact results.  The algorithm would, however, benefit from tools that would allow one to estimate the error.  At least in the context of symmetry groups we are optimistic that this may be done.  If a choice of reference state can be made for which the complexity is analytically computable, a more general choice can be made by performing a unitary transformation of either the state or Hamiltonian (but not both).  It may be possible to estimate the error by imposing a suitable metric on the space of such unitary transformations.  \\ \\
Finally, in this paper we have focused on the example of a time-independent Hamiltonian.  Extending to Hamiltonians that carry an explicit time-dependence would be an important generalisation of this modified Lanczos method.

\acknowledgments

We would like to thank Pawel Caputa, Bret Underwood and Arpan Bhattacharyya for discussions and comments. JM would like to acknowledge support from the ICTP through the Associates Programme and from the Simons Foundation through grant number 284558FY19. SH would like to thank Brac University for their hospitality during the final stage of this work. H.J.R.vZ is supported by the ``Quantum Technologies for Sustainable Devlopment" grant from the National Institute for Theoretical and Computational Sciences (NITHECS).

\appendix

\section{Special choice of reference state}
\label{refStateAppendix}

In this appendix we would like to highlight a choice of reference state for which the Krylov complexity is exactly solvable.  This choice of reference state is dependent on the coefficients appearing in the Hamiltonian and therefore, obviously a special choice.  Nevertheless, it is useful to recognize that there exists reference states for which the problem is solvable. 
We first observe that we may identify various $su(1,1)$ subalgebras of the Jacobi algebra
\begin{eqnarray}
\tilde{L}_{+} & = & L_{+} + c l_{+} + \frac{c^2}{2} l_0\,,   \nonumber \\
\tilde{L}_{-} & - & L_{-} + c^* l_{-} + \frac{(c^*)^2}{2} l_0\,,     \\
\tilde{L}_0 & = & L_0 + \frac{c^*}{2} l_{+} + \frac{c}{2} l_{-} + \frac{c c^*}{2} l_0\,.\nonumber
\end{eqnarray}
The Hamiltonian (\ref{JacobiH}) may be written (up to an overall central element) as a linear combination of these generators
\begin{equation}
H = \alpha \tilde{L}_{+} + \alpha^* \tilde{L}_{-} + \gamma \tilde{L}_0 - \left( \alpha \frac{c^2}{2} + \alpha^* \frac{(c^*)^2}{2} + \gamma \frac{c c^*}{2}   \right) l_0\,,
\end{equation}
with the choice
\begin{equation}
c = \frac{4 \alpha^* \beta - 2 \beta^* \gamma}{4 |\alpha|^2 - \gamma^2}\,.
\end{equation}
If one can find a reference state such that
\begin{eqnarray}
\tilde{L}_{-} |\phi_r\rangle & = & 0 \nonumber \\
\tilde{L}_0 |\phi_r\rangle  & = & \Delta |\phi_r\rangle
\end{eqnarray}
the problem reduces to one of $SU(1,1)$ coherent states, for which exact analytic results are known \cite{Caputa:2021sib}.  This choice of reference state is given by 
\begin{equation}
|\phi_r\rangle = e^{-c^* l_{+} }|h, 0, M\rangle
\end{equation}
for which it can be shown without too much trouble that $\Delta = 0$.  The exact Krylov basis is given by
\begin{equation}
|K_n\rangle = \mathcal{N}_n (\tilde{L}_{+} )^{2n} e^{-c^* l_{+} }|h, 0, M \rangle.  
\end{equation}
Note that this choice of reference state for the Hamiltonian is unitarily equivalent to setting $\beta \rightarrow 0$ with the reference state $|h, 0, M\rangle$ we have considered in the main text.

\section{Nielsen complexity for Jacobi coherent states}

\begin{figure}
    \centering
    \hspace{-1.0cm}
    \includegraphics[width=0.70\textwidth]{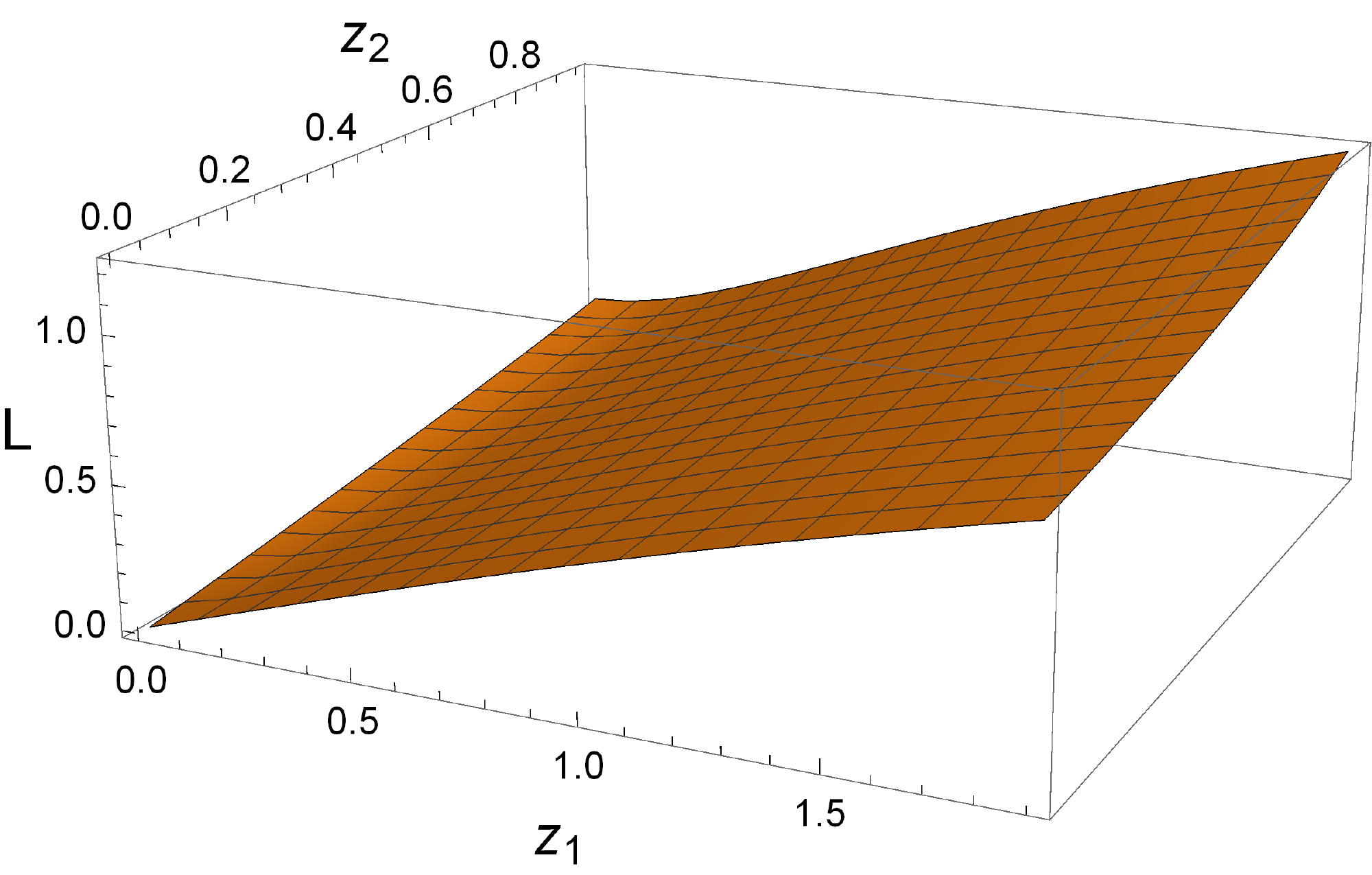}
    \caption{The geodesic length connecting the origin to the point $z_1, z_2$ on the manifold with both $z_1$ and $z_2$ real. 
 We have set $h=\frac{1}{4}$ and $M=1$.  Note that the complexity increases with increasing $|z_i|$ but that the gradient is different for the two directions.}
    \label{fig:NielsenComp1}
\end{figure}

\begin{figure}
    \centering
    \hspace{-1.0cm}
    \includegraphics[width=0.70\textwidth]{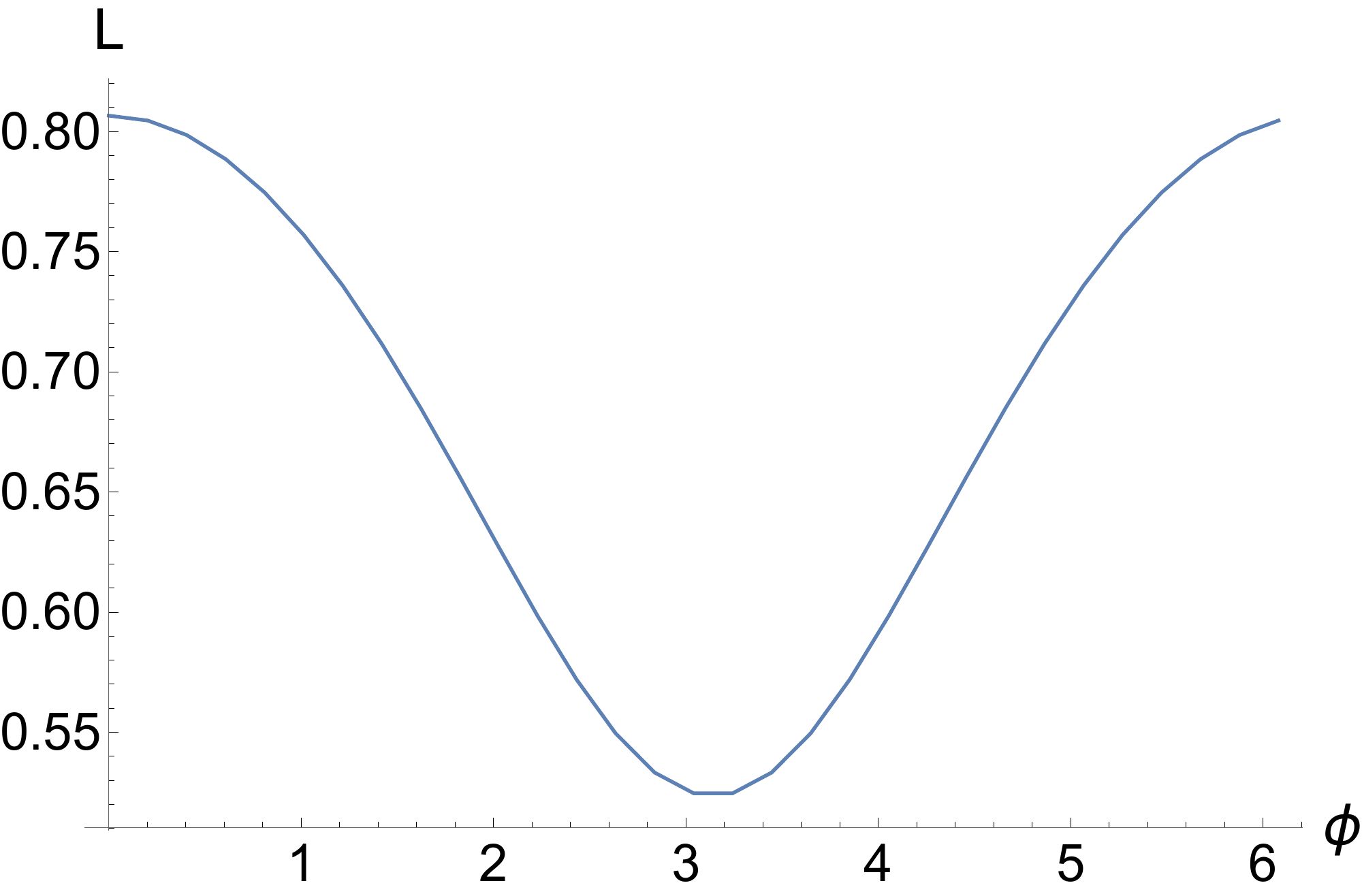}
    \caption{The geodesic length connecting the origin to the point $z_1, z_2$ on the manifold with $z_1  = 2$ and $z_2 = 0.8 e^{i \phi}$.  We have set $h=\frac{1}{4}$ and $M=1$. }
    \label{fig:NielsenComp2}
\end{figure}

In this appendix we unpack some results for the Nielsen complexity of Jacobi coherent states.  Briefly, for Nielsen complexity one needs to specify a set of continuous unitary gates, a reference state, target state and a ``cost'' function on the space of unitary gates.  Our choice of unitary gates is taken as the Jacobi symmetry group.  The accessible target states are those that differ from the reference state by a Jacobi unitary transformation and are, in fact, the Jacobi group coherent states.  For the lowest weight reference state $|h, 0, M\rangle$ we have
\begin{equation}
|z_1, z_2\rangle = N(z_1, z_2) e^{z_1 l_{+}} e^{z_2 L_{+}} |h, 0, M\rangle.  
\end{equation}
As the cost function we will use the Fubini-Study metric, which may be obtained from the coherent states above as 
\begin{equation}
g_{a \bar{b}} = \partial_{z_a} \partial_{\bar{z}_{b}} \log ( \langle h, 0, M| e^{\bar{z}_1 l_{-}} e^{\bar{z}_2 L_{-}} e^{z_1 l_{+}} e^{z_2 L_{+}}   |h, 0, M\rangle     ). 
\end{equation}
We have already computed the coherent state overlap in the main text in eq(\ref{JacobiCSOL}). Consequently, the Fubini-Study metric is easily computed as
\begin{eqnarray}
ds^2 & = & \frac{M}{1 - |z_2|^2} dz_1 d\bar{z}_1 +  \frac{M | z_1 + \bar{z}_1 z_2|^2 + 2 h (1 - |z_2|^2)}{(1 - |z_2|^2   )^3} dz_2 d\bar{z}_2 \nonumber  \\
&+& \frac{M(z_1 + \bar{z}_1 z_2)}{2(1 - |z_2|^2 )^2} dz_1 d\bar{z}_2 +  \frac{M(\bar{z}_1 + z_1 \bar{z}_2)}{2(1 - |z_2|^2 )^2} d\bar{z}_1 d z_2\,.
\end{eqnarray}
The (Nielsen) complexity may now be computed as the minimal geodesic distance connecting the reference and target states.  On the $z_1 = 0$ submanifold the geodesic may be solved as
\begin{equation}
z_2(\sigma) = e^{i \phi} \tanh(c \sigma)   
\end{equation}
recovering the results of (\cite{Chapman:2021jbh}) for the geodesic distance when plugging this into 
\begin{equation}
L = \int_{\sigma_{min}}^{\sigma_{max}} \sqrt{ \left| \sum_{a,b}g_{\bar{a} b} \bar{z}'_a(\sigma) z'_b(\sigma) 
  \right|} \ d\sigma
\end{equation}
For general points on the manifold the geodesic equation may be solved numerically.  For real $z_1, z_2$ we find that the geodesic length (i.e. complexity) is a monotonically increasing function, see Fig (\ref{fig:NielsenComp1}).  The gradient for the geodesic length has a different slope in the $z_1$, $z_2$ directions.  
\\ \\
 Finally, in the same way that we noted a dependence on the phases $a,b$ in the context of spread complexity, the Nielsen complexity also displays a dependence on the phase of $z_i$, see Fig. (\ref{fig:NielsenComp2})

\section{Lanczos Algorithm for our Hamiltonian}

\label{LanczosAppendix}

In this appendix we demonstrate some of the involved combinatorics that arise when the Lanczos algorithm is applied to the Jacobi Hamiltonian and highest weight state considered in the main text.  By definition we have as starting point
\begin{equation}
|K_0\rangle = |h,0,M\rangle
\end{equation}
where we set $h=\frac{1}{4}$.  Acting on this state with the Hamiltonian gives
\begin{equation}
H |K_0\rangle = \frac{\alpha}{\sqrt{2}} |h, 1, M\rangle + \sqrt{M} \beta |h, \frac{1}{2}, M\rangle + \frac{\gamma}{4} |h, 0, M\rangle
\end{equation}
from which we obtain
\begin{eqnarray}
|A_1) & = & \frac{\alpha}{\sqrt{2}} |h, 1, M\rangle + \sqrt{M} \beta |h, \frac{1}{2}, M\rangle   \nonumber \\
b_1 & = & \frac{1}{\sqrt{2}}\sqrt{  |\alpha|^2 + 2 M |\beta|^2}\,,
\end{eqnarray}
and $|K_1\rangle = \frac{1}{b_1} |A_1)$.  The next step in the Lanczos algorithm then yields
\begin{eqnarray}
H|K_1\rangle & = & \frac{1}{\sqrt{2}}\sqrt{|\alpha|^2 + 2 M |\beta|^2 } |h, 0, M\rangle  +   \frac{\sqrt{M}(4 \alpha \beta^* + 3 \beta \gamma)   }{2 \sqrt{2} \sqrt{|\alpha|^2 + 2 M |\beta|^2 }}|h, \frac{1}{2}, M\rangle + \frac{8 M \beta^2 + 5 \alpha \gamma}{4 \sqrt{|\alpha|^2 + 2 M |\beta|^2}} |h, 1, M\rangle      \nonumber \\
&+& \frac{2 \sqrt{3} \sqrt{M} \alpha \beta}{\sqrt{|\alpha|^2 + 2 M |\beta|^2}} |h, \frac{3}{2}, M\rangle + \frac{\sqrt{3} \alpha^2}{\sqrt{|\alpha|^2 + 2 M |\beta|^2}}|h, 2, M\rangle\,,
\end{eqnarray}
from which we extract
\begin{eqnarray}
|A_2) & = & \frac{4 M \beta^3 - 2 \alpha^2 \beta^* + \alpha \beta \gamma }{\sqrt{2} (|\alpha|^2 + 2 M |\beta|^2)^{\frac{3}{2} } }\left( 2 M \beta^* |h, 1, M\rangle -\sqrt{2} \sqrt{M} \alpha^* |h, \frac{1}{2}, M\rangle   \right)    \nonumber \\
&+& \frac{\sqrt{3} \alpha}{\sqrt{|\alpha|^2 + 2 M |\beta|^2}}\left( \alpha |h, 2, M\rangle + 2\sqrt{M} \beta |h, \frac{3}{2}, M\rangle   \right)\,,    \nonumber \\
b_2 & = & \left( \frac{3|\alpha|^2(  |\alpha|^2 + 4 M |\beta|^2 )}{|\alpha|^2 + 2 M |\beta|^2}  +   \frac{M |4 M \beta^3 - 2\alpha^2 \beta^* + \alpha \beta \gamma|^2  }{2 ( |\alpha|^2 + 2 M |\beta|^2   )^2}     \right)^{\frac{1}{2}}\,,
\end{eqnarray}
and $|K_2\rangle = \frac{1}{b_2} |A_2)$.  Note that this vector, compared to $|K_1\rangle$ contains both a linear combination of $|h, \frac{3}{2}, M\rangle$, $|h, 2, M\rangle$ as well as the linear combination of $|h, \frac{1}{2}, M\rangle$, $|h, 1, M\rangle$ orthogonal to $|K_1\rangle$.  Similarly, the next Krylov basis vector will contain a linear combination of $|h, \frac{5}{2}, M\rangle$, $|h, 3, M\rangle$ as well as a vector orthogonal to $|K_2\rangle$ and $|K_1\rangle$.  The specific linear combinations of vectors become progressively more unwieldy as the algorithm progresses.

\bibliographystyle{JHEP}
\bibliography{kitaevchainrefs.bib}

\end{document}